# Spin nematic liquid crystal and scalar spin chirality in tetragonal lattice $YbMnBi_2$

Yaofeng Xie[1,2*], Sijie Xu[1,2*], Yu Pan[3,4,*], Taekoo Oh[5,6,10*], Tingjun Zhang[1,2,11*], Masaaki Matsuda[7], Zhaoyu Liu[1,2], Zehao Wang[1,2], Yiheng Wang[1,2], Siyu Pan[1,2], Avishek Maity[7], Sylwia Pawledzio[7], Xiaoping Wang[7], Songxue Chi[7], Feng Ye[7], Yiqing Hao[7], Huibo Cao[7], Barry L. Winn[7], Melissa K. Graves-Brook[7], Shuai Wu[8], Fan Li[8], Xiaoyuan Zhou[8], Claudia Felser[4], Naoto Nagaosa[6,9] and Pengcheng Dai[1,2]

[1] Department of Physics and Astronomy, Rice University, Houston, Texas 77005, USA

[2]Rice Laboratory for Emergent Magnetic Materials and Smalley-Curl Institute, Rice University, Houston, Texas 77005, USA

[3]College of Materials Science and Engineering & Center of Quantum Materials and Devices, Chongqing University, Chongqing 401331, People's Republic of China

[4]Max Planck Institute for Chemical Physics of Solids, Dresden 01187, Germany

[5]Department of Physics, Soongsil University, Seoul 06978, South Korea

[6]RIKEN Center for Emergent Matter Science (CEMS), Wako, Saitama, 351-0198 Japan

[7]Neutron Scattering Division, Oak Ridge National Laboratory, Oak Ridge, Tennessee 37831, USA

[8]College of Physics, Chongqing University, Chongqing 401331, People's Republic of China

[9]Fundamental Quantum Science Program (FQSP), TRIP Headquarters, RIKEN, Wako, Saitama, 351-0198, Japan

[10]Origin of Matter and Evolution of Galaxies (OMEG) Institute, Soongsil University, Seoul 06978, South Korea

[11]Applied Physics Graduate Program, Smalley-Curl Institute, Rice University, Houston, Texas 77005, USA

*These authors made equal contributions to the projects.

Correspondence should be addressed to N.N. (nagaosa@riken.jp) or P.D. (pdai@rice.edu).

**A spin nematic order, analogous to the nematic liquid crystal, characterizes the spontaneous breaking of spin-space rotational symmetry while preserving time-reversal (*T*) symmetry. On the other hand, the order parameter characterizing the *T*-symmetry breaking is the composite three spin order, i.e., the scalar spin chirality (SSC) $\chi_{ijk} = \langle \boldsymbol{S_i} \cdot (\boldsymbol{S_j} \times \boldsymbol{S_k}) \rangle$, where $\boldsymbol{S_i}, \boldsymbol{S_j}, \boldsymbol{S_k}$ are spins at neighboring sites $\boldsymbol{i}, \boldsymbol{j}, \boldsymbol{k}$, respectively, and nonzero SSC is known to induce anomalous Hall effect (AHE). Although a spin nematic phase has been suggested in the frustrated magnets and the square-lattice iridate, how a spin nematic phase might affect magneto-transport properties is unknown. Here we use polarized neutron scattering to show that tetragonal lattice *A*$MnBi_2$ (*A*=Ca, Yb) is strictly a *c*-axis aligned collinear antiferromagnet (C-type) with $T_N \approx 270$ K and 290 K. On cooling from 450 K to $T_N$, low-energy spin excitations in $YbMnBi_2$ spontaneously change from isotropic to anisotropic in spin space within the tetragonal plane, forming a dynamic spin nematic phase around 400 K due to heavy Yb-induced spin-orbit coupling, before gapping out below $T_N$. Similar polarized neutron scattering measurements on $CaMnBi_2$ reveal isotropic paramagnetic scattering without a spin nematic phase above $T_N$. Under an in-plane external magnetic field, the $Yb^{3+}$ moments may interact with the dynamic spin nematic phase to induce nonzero SSC,**

**giving rise to AHE and anomalous Nernst effect (ANE) in $YbMnBi_2$ that is absent in $CaMnBi_2$ above $T_N$. Theoretical analysis of Ginzburg-Landau theory based on the symmetry indicates that the coupling terms between the nematic order and SSC of 5th order in the spin operators are allowed under an external magnetic field $B$. This could explain the rapid increase of AHE as a function of B in $YbMnBi_2$. Our results, therefore, provide compelling evidence for dynamic SSC-induced AHE and ANE in the paramagnetic phase of a compensated collinear antiferromagnet, opening a new avenue for the physics of composite orders of multiple spins for room-temperature spintronics without magnetic order.**

## I. Introduction

Long-range orders are the most important subject in the physics of magnets. A Hamiltonian that describes these systems is the Heisenberg model (1, 2), where the magnetic ordering temperature is determined by the competition between the thermal energy and exchange interaction $H = -J\boldsymbol{S}_i \cdot \boldsymbol{S}_j$ ($J$ is the magnetic exchange coupling) of the two neighboring spins $\boldsymbol{S}_i$ and $\boldsymbol{S}_j$, which respects the spin rotational symmetry. In simple ferromagnets, all the spins $\langle \boldsymbol{S}_i \rangle$ align along the same direction, where the order parameter is characterized by a finite single-spin expectation value $\langle \boldsymbol{S}_i \rangle$, which breaks general spin rotation symmetry together with the time-reversal symmetry. Although antiferromagnets have alternating $\langle \boldsymbol{S}_i \rangle$ and $\langle \boldsymbol{S}_j \rangle$ values, the order parameter is still described by a linear combination of single-spin order parameters.

Beyond the single-spin order parameters, there are spin-related orders composed of multiple spins, such as the spin Peierls order, where the dimers of spins are formed and characterized by the order parameter $<\boldsymbol{S}_i \cdot \boldsymbol{S}_j>$, which breaks the translational symmetry while the time-reversal and spin rotation symmetries are intact (3). Another multiple spin-related order is the spin nematic state where the spin rotational symmetry is broken not by a single-spin order ($\langle \boldsymbol{S}_i \rangle = 0$), but by an anisotropic two-spin correlation like $\langle S_i^a S_j^b \rangle$ where $a, b = x, y, z$ are along different crystallographic directions (4, 5). This spin nematic order is characterized by a broken rotational symmetry in spin space without breaking the time reversal ($T$-) symmetry, in contrast to the electronic nematic phase in quantum materials that breaks the $C_4$ rotational symmetry of the underlying lattice (6-9). It can emerge from four-spin exchange interactions (10), quadrupolar order (5), magnetic frustrations (11, 12), or spin-orbit coupling (SOC) (13-15), and can occur in two-dimensional (2D) tetragonal lattice materials with the $C_4$ rotational symmetry (13-18). On the other hand, the multiple-spin-related order parameter characterizing the $T$-symmetry breaking is the scalar spin chirality (SSC) $\chi_{ijk} = \langle \boldsymbol{S}_i \cdot (\boldsymbol{S}_j \times \boldsymbol{S}_k) \rangle$, a three-spin correlation composite order (19-21). It is well known that SSC acts as the emergent magnetic field and induces anomalous Hall effect (AHE) (Fig. 1a) (22, 23), and can arise from the imaginary part of four-spin exchange interactions. In general, intrinsic AHE can be induced by several different mechanism including: (1) the long-range ferromagnetic order (23); (2) the real space noncoplanar spin texture (skyrmions) with finite SSC in helimagnets (20, 21, 24) and noncollinear antiferromagnets (25); (3) the SOC induced Berry curvature of filled electronic bands in nonmagnetic (26-29) and magnetic (30-32) Weyl semimetals; and (4) a compensated collinear antiferromagnet with a magnetic toroidal quadrupole Fermi surface (an altermagnet) when SOC is included (33, 34) (Fig. 1a). Therefore, AHE can appear deep inside the magnetically ordered phase or near a magnetic phase transition, and SSC can appear with either the noncoplanar spin texture or the spin fluctuations in the absence of magnetic ordering (35). Since a spin nematic phase does not break the $T$-symmetry and can appear in a paramagnetic phase well above the magnetic transition temperature (4, 5, 11-15), one should consider the link of the spin nematic phase to the SSC from the spin fluctuation to induce AHE.

$YbMnBi_2$, known to crystallize in a tetragonal structure with a C-type antiferromagnetic (AFM) structure (Fig. 1b) and the space group $P4/nmm$ with inversion lattice symmetry (36-39), is a perfect platform to investigate the spin nematic phase due to heavy Bi and Yb-induced large SOC and its connection with AHE through the SSC for the following reasons. Previously, AHE/ANE over a wide temperature range around $T_N$ was reported and attributed to the Berry curvature effect induced by a small canted moment away from the *c*-axis aligned AFM order (40-43). According to density functional theory (DFT) calculations, finite Dzyloshinskii-Moriya (DM) interactions in the Mn-Bi-Mn bonds arising from Bi vacancies in $YbMnBi_2$ can cant the *c*-axis moment in the collinear C-type antiferromagnet (43). When the canting angle is greater than 2$^{o}$, the net ferromagnetism from canted Mn moment can give rise to a *T*-symmetry-breaking type-II Weyl state and therefore the observed AHE/ANE (43). Angle-resolved photoemission spectroscopy (ARPES) experiments on $YbMnBi_2$ reported signatures of a *T*-symmetry-breaking type-II Weyl semimetal and interpreted the results within the same canted-AFM framework (Fig. 1b) (44). However, unpolarized and polarized neutron scattering experiments find no evidence of an observable canted moment (36-39). Since unpolarized and polarized neutron diffraction experiments find that $CaMnBi_2$ has an identical AFM structure to that of $YbMnBi_2$ (37, 38, 45-47), it will be interesting to carry out AHE measurements on $CaMnBi_2$ and compare with $YbMnBi_2$ under a magnetic field along the a-axis over a wide temperature range around AFM order (Figs. 1c-1e). Surprisingly, the values of AHE in $CaMnBi_2$ determined from our AHE measurements are one order of magnitude smaller than those of $YbMnBi_2$ (Figs. 1f and 1g), indicating significant differences in the microscopic mechanism responsible for the magneto-transport properties of these two materials. Therefore, it is essential to thoroughly investigate the magnetic properties of these materials and determine whether they exhibit a spin nematic phase and whether AHE is related to SSC.

Here we report the discovery of a spontaneous high-temperature spin nematic phase ($T_S \gg T_N$) in $YbMnBi_2$ and its coupling to AHE/ANE via field-induced SSC. Using neutron longitudinal polarization analysis (LPA) (2, 8, 48), we can conclusively measure the magnetic responses (elastic and inelastic) of $A MnBi_2$ along the *a*-, *b*-, and *c*-axis directions denoted as $M_a$, $M_b$, and $M_c$ (Figs. 1a and 1b). For both $CaMnBi_2$ and $YbMnBi_2$, the magnetic structure is a strictly a *c*-axis aligned collinear C-type AFM state, with no evidence for an in-plane canting angle larger than 0.3º (37, 38, 45-47) or for a ferrimagnetic component, contrary to the previous reports (40-43). Because neutron scattering is a bulk probe not sensitive to surface effects, these measurements rule out the previously proposed bulk canting-induced type-II Weyl scenario for $A MnBi_2$ (28, 44). While $YbMnBi_2$ remains tetragonal over temperature range probed (37, 38), recent polarized neutron scattering and X-ray diffraction experiments find that $CaMnBi_2$ undergoes an orthorhombic lattice distortion below ~46 K with unit-cell doubling along the *c* axis due to a Peierls transition, with minimal impact on the magnetic order (47). On cooling toward $T_N$, low-energy ($E < 4$ meV) spin excitations of $YbMnBi_2$ spontaneously change from isotropic ($M_a \approx M_b \approx M_c$) to anisotropic ($M_a \approx M_c > M_b$) in spin space around $T_S \approx 400$ K, forming a dynamic spin nematic phase arising from the Yb-induced SOC (49-51) before gradually gapping out below $T_N$. In contrast, analogous polarized neutron scattering measurements on $CaMnBi_2$ reveal isotropic paramagnetic scattering without a spin nematic phase above $T_N$. Our observation of a spin nematic phase in $YbMnBi_2$ is unique because it directly resolves the magnitude of spin correlations in different crystallographic directions, which cannot be achieved with the $\boldsymbol{Q} = 0$ probe such as Raman scattering (16). X-ray photoelectron spectroscopy (XPS) shows that more than half of the Yb ions in $YbMnBi_2$ exist as $Yb^{3+}$ and carry magnetic moments, whereas $Ca^{2+}$ remains non-magnetic. However, the $Yb^{3+}$ ions in $YbMnBi_2$ do not behave as nearly free, weakly interacting local moments. Isotropic free $Yb^{3+}$ ion has $J = 7/2$, but in a crystal field of $YbMnBi_2$, $J = 7/2$ is split into four Kramers doublets with an effective pseudospin-1/2 ground state (52-55). As shown below, high-energy inelastic neutron scattering reveals Yb3+ crystal-electric-field (CEF) excitations at 92.2, 163.7, and 180.2 meV, and the corresponding CEF analysis yields an anisotropic ground-state Kramers doublet with a strongly reduced *c* axis dipolar moment. Together with Yb-Mn coupling and possible valence fluctuations/hybridization (52-54), this explains why the bulk susceptibility is much smaller than expected for independent free $Yb^{3+}$ moments. Under an in-plane magnetic field $\boldsymbol{B}$, these field-polarizable $Yb^{3+}$ degrees of freedom can couple to the dynamic spin nematic phase and induce dynamic SSC, giving rise to AHE (23) and ANE (59) by breaking *T*-symmetry (40, 41). Our results therefore provide compelling evidence for a close link between two higher-order composite spin orders, spin nematicity and dynamic SSC, in the paramagnetic phase of a compensated collinear antiferromagnet, pointing to a route toward room-temperature spintronic responses without relying on ferromagnetism or spin canting.

## II. Results

### A. Experimental Data

We use neutron LPA to elucidate the magnetic structure of $A$MnBi$_2$ and understand the contrasting anomalous transport phenomena in these compounds (Figs. 1f, 1g). Since neutron scattering can only measure magnetic moments perpendicular to momentum transfer $\boldsymbol{Q}$ (2) and $A$MnBi$_2$ has the C-type AFM order with dominant moment along the $c$-axis (Fig. 1b, $M_c$) (37, 38, 45-47), we study the $c$-axis and $ab$-plane magnetic moments by probing magnetic Bragg peak positions and $\boldsymbol{Q} = (0,0,L)$, where $(H,K,L)$ are Miller indices, respectively. We adopt a conventional coordinate system for polarized neutron studies with the $\boldsymbol{x}$-axis along $\boldsymbol{Q}$, $\boldsymbol{y}$-axis orthogonal to $\boldsymbol{Q}$ within the $(H,0,L)$ scattering plane, and $\boldsymbol{z}$-axis perpendicular to the scattering plane (Fig. 2a) (8, 48). By aligning the incident neutron polarization to be parallel to the $\boldsymbol{x}$, $\boldsymbol{y}$, or $\boldsymbol{z}$ axis, and the scattered neutrons to be parallel or antiparallel to the $\boldsymbol{x}$, $\boldsymbol{y}$, or $\boldsymbol{z}$ axis, we can measure neutron spin-flip (SF) cross sections $\sigma_{x,y,z}^{\mathrm{SF}}$ and non-spin-flip (NSF) cross sections $\sigma_{x,y,z}^{\mathrm{NSF}}$ (Figs. 2a-2c). Here, we focus on four cross sections that are most important to our study: $\sigma_x^{\mathrm{NSF}} \propto NC$, $\sigma_x^{\mathrm{SF}} \propto M_y + M_z$, $\sigma_y^{\mathrm{SF}} \propto M_z$, and $\sigma_z^{\mathrm{SF}} \propto M_y$, where $NC$ is the coherent nuclear Bragg scattering, $M_y$ and $M_z$ are magnetic scattering along the $y$ and $z$ directions, respectively. We can further use $M_y \propto \sigma_x^{\mathrm{SF}} - \sigma_y^{\mathrm{SF}}$ and $M_z \propto \sigma_x^{\mathrm{SF}} - \sigma_z^{\mathrm{SF}}$ to separate magnetic scattering $M_y$ and $M_z$ from nuclear and background contributions. Neutron LPA thus provides a conclusive way to determine the $c$-axis and $ab$-plane magnetic responses in $A$MnBi$_2$ (Fig. 1b).

To confirm the C-type AFM order and determine whether there is a magnetic moment within the $ab$ plane as suggested by ARPES (44), we measured $\sigma_x^{\mathrm{NSF}}$ and $\sigma_{x,y,z}^{\mathrm{SF}}$ around $\boldsymbol{Q} = (0,0,1)$, (1,0,0), and (2,0,0) on $A$MnBi$_2$ (Figs. 2d-2h). Figure 2d shows the $L$-scan of NSF $\sigma_x^{\mathrm{NSF}}/10$ and SF cross sections $\sigma_{x,y,z}^{\mathrm{SF}}$ across $\boldsymbol{Q} = (0,0,1)$ on YbMnBi$_2$ at 50 K. The dominant signal at $\boldsymbol{Q} = (0,0,1)$ is the structural Bragg scattering, as shown in the large intensity of $\sigma_x^{\mathrm{NSF}}$. The small peak at $\boldsymbol{Q} = (0,0,1)$ in $\sigma_{x,y,z}^{\mathrm{SF}}$ is due to finite neutron polarization ratio (see methods). Since all SF channels exhibit identical intensity ($\sigma_x^{\mathrm{SF}} = \sigma_y^{\mathrm{SF}} = \sigma_z^{\mathrm{SF}}$), there is no detectable magnetic scattering anywhere perpendicular to the *c*-axis, suggesting that the magnetic moment lies strictly along the *c*-axis. Figures 2e and 2f show in-plane scans of $\sigma_{x,y,z}^{\mathrm{SF}}$ across $\boldsymbol{Q} = (1,0,0)$, and (2,0,0) at 50 K. The peaks in $\sigma_x^{\mathrm{SF}} - \sigma_y^{\mathrm{SF}}$ at $\boldsymbol{Q} = (1,0,0)$ corresponds to the *c*-axis AFM moment as expected, the featureless $\sigma_x^{\mathrm{SF}} - \sigma_z^{\mathrm{SF}}$ suggests that there is no in-plane AFM moment along the (0,1,0) direction (Fig. 2e, methods). Figure 2f shows similar measurements at $\boldsymbol{Q} = (2,0,0)$, where the C-type AFM order has no contribution and therefore will be ideal to probe the potential *c*-axis ferrimagnetism. We find negligible differences between $\sigma_x^{\mathrm{SF}}$ and $\sigma_y^{\mathrm{SF}}$, indicating that the *c*-axis moments remain strictly AFM without ferrimagnetic component (Fig. 2f). Again, $\sigma_x^{\mathrm{SF}}$ and $\sigma_z^{\mathrm{SF}}$ at $\boldsymbol{Q} = (2,0,0)$ are of similar magnitude, confirming the absence of an in-plane net moment along the (0,1,0) direction (Fig. 2f).

To compare with $YbMnBi_2$, we performed the same measurements across $\boldsymbol{Q} = (0,0,1)$, (1,0,0) on $CaMnBi_2$. As shown in Fig. 2g, the *L*-scans again reveal $\sigma_x^{\mathrm{SF}} = \sigma_y^{\mathrm{SF}} = \sigma_z^{\mathrm{SF}}$, indicating the absence of in-plane magnetic moments. The difference between $\sigma_x^{\mathrm{SF}}$ and $\sigma_y^{\mathrm{SF}}$ of *H*-scan at (1,0,0) reflects the *c*-axis AFM moments. Figure 2i shows the temperature dependence of the integrated peak intensity of $\sigma_x^{\mathrm{SF}} - \sigma_y^{\mathrm{SF}}$ at $\boldsymbol{Q} = (1,0,0)$ for both compounds. A power-law fit to the temperature dependence of the peak intensity reveals a Néel temperature $T_N \approx 290$ K for $YbMnBi_2$ and 270 K for $CaMnBi_2$. These results establish that both $YbMnBi_2$ and $CaMnBi_2$ exhibit strictly collinear C-type AFM order with magnetic moments confined to the *c*-axis, and no evidence of in-plane canting to within ~0.3° or ferrimagnetic component (see methods), consistent with recent polarized neutron scattering results (47). Since a minimum of 2° canting angle is required to drive the system into the type-II Weyl state (43), our measurements rule out the previously proposed canting-induced bulk type-II Weyl scenario for *A*$MnBi_2$ (44).

To search for spin space anisotropy in $YbMnBi_2$, we carried out inelastic neutron scattering experiments using the LPA to measure $\sigma_{x,y,z}^{\mathrm{SF}}$ and $\sigma_x^{\mathrm{NSF}}$ at the AFM zone centers $\boldsymbol{Q}_1 = (1,0,0)$, $\boldsymbol{Q}_2 = (1,0,1)$, and $\boldsymbol{Q}_3 = (1,0,2)$ (Figs. 3a-3l). By measuring $\sigma_{x,y,z}^{\mathrm{SF}}$ at two of the three $\boldsymbol{Q}$s, we can conclusively determine the magnetic responses of $YbMnBi_2$ along the *a*-, *b*-, *c*-axis directions $(M_a, M_b, M_c)$ as a function of temperature across $T_N$ (Figs. 1a, 3m-3p, 5e-5g) (49, 55). Figures 3a and 3e show the energy scans of neutron SF scattering cross sections $\sigma_x^{\mathrm{SF}}$, $\sigma_y^{\mathrm{SF}}$, and $\sigma_z^{\mathrm{SF}}$ at $\boldsymbol{Q}_1$ and $\boldsymbol{Q}_3$, respectively, at $T$ = 230 K ($< T_N$). We find a spin gap of about ~7 meV, arising from SOC-induced uniaxial anisotropy that determines the easy axis of the magnetic ordered moment, in all three SF $\sigma_{x,y,z}^{\mathrm{SF}}$ channels, consistent with earlier unpolarized inelastic neutron scattering work (37, 38). Above the spin gap energy, spin waves at $\boldsymbol{Q}_1$ have $\sigma_x^{\mathrm{SF}} \approx \sigma_y^{\mathrm{SF}}$ and $\sigma_z^{\mathrm{SF}} \approx 0$ (Fig. 3a), thus indicating that they are entirely transverse and perpendicular to the *c*-axis ordered moment as expected from a local-moment Heisenberg model ($M_a \approx M_b, M_c \approx 0$, Fig. 3i) (37, 38).

Figures 3b and 3f show identical $\sigma_{x,y,z}^{\mathrm{SF}}$ measurements at $T$ = 300 K ($\approx T_N + 10$ K) in the paramagnetic state, where the spin gap collapses (37, 38). For excitation energies above 4 meV, we have $\sigma_x^{\mathrm{SF}} > \sigma_y^{\mathrm{SF}} \approx \sigma_z^{\mathrm{SF}}$, consistent with isotropic paramagnetic excitations as expected. For $E < 4$ meV, we find $\sigma_x^{\mathrm{SF}} > \sigma_z^{\mathrm{SF}} > \sigma_y^{\mathrm{SF}}$, suggesting the presence of spin excitation anisotropy (49). Since magnetic exchange couplings within the *ab*-plane are much larger than those along the *c*-axis (37, 38), we expect paramagnetic spin excitations to be consistent with a 2D Heisenberg model with weak interlayer coupling. On cooling to $T_N$, spin excitations in $YbMnBi_2$ should have a 2D to 3D crossover where the longitudinal component $M_c$ is expected to diverge at $T_N$, while the transverse components $M_a, M_b$ should remain noncritical with $M_a = M_b$ at $T_N$, similar to spin excitations in AFM ordered $BaFe_2As_2$ (49-51, 56). However, clear anisotropic scattering below $E \approx 4$ meV with $M_c \approx M_a > M_b$ (Fig. 3j) implies the formation of in-plane spin space anisotropy and a dynamic spin nematic phase above $T_N$.

At 400 K ($T_s \approx T_N + 110$ K), $\sigma_x^{\mathrm{SF}}$ shows a broad peak at ~7 meV, and $\sigma_y^{\mathrm{SF}}$ becomes the same as $\sigma_z^{\mathrm{SF}}$ suggesting isotropic paramagnetic scattering at all energies (Figs. 3c and 3g). The energy dependence of $M_a, M_b, M_c$, obtained using $\sigma_{x,y,z}^{\mathrm{SF}}$ at $\boldsymbol{Q}_1, \boldsymbol{Q}_3$ (Fig. 3m) (55), shows a broad peak around ~7 meV with $M_a \approx M_b \approx M_c$ (Fig. 3k). At $T$ = 450 K ($\approx T_N + 160$ K), the results remain consistent with those at 400 K, showing isotropic magnetic fluctuations across the measured energy range and a broad feature centered around 3-7 meV (Figs. 3d, 3h, and 3l). Figure 3n shows the temperature dependence of $M_a, M_b, M_c$, and Figure 3o shows the *ab*-plane magnetic anisotropy, defined as $M_a - M_b$, at $E \approx 2$ meV. At 450 K, spin fluctuations are isotropic with $M_a \approx M_b \approx M_c$. As temperature decreases towards $T_N$, magnetic anisotropy spontaneously emerges below $T_S$ with $M_a - M_b$ peaking at $T_N$ (Figs. 3n and 3o). Figure 3p shows the temperature evolution of the spin excitations $M_a, M_b, M_c$ and $M_a - M_b$ at $E \approx 9$ meV. While spin excitations are isotropic with $M_a \approx M_b \approx M_c$ above $T_N$, they become transverse spin waves below $T_N$ with $M_a \approx M_b, M_c = 0$ as expected.

To understand details of the dynamic spin nematic phase in $YbMnBi_2$, we show $E$ and $\boldsymbol{Q}$ dependent scattering probed by unpolarized neutrons along the in-plane $[H, 0,0]$ (Figs. 4a-4d) and *c*-axis $[1,0, L]$ (Figs. 4e-4h) directions across the broad temperature range in Fig. 3. At 270 K, we see well defined magnetic scattering at all probed energies along the $[H, 0,0]$ (Fig. 4a) and $[1,0, L]$ (Fig. 4e) directions, although a spin gap starts to form around 5 meV. On warming to above $T_N$ at 300 K and 350 K, spin excitations at probed energies are still well-defined along the $[H, 0,0]$ (Figs. 4b and 4c) and $[1,0, L]$ (Figs. 4f and 4g) directions. While the in-plane spin excitations are still well defined around [1,0,0] at 450 K (Fig. 4d), spin-spin correlations along the *c*-axis disappear at all probed energies (Fig. 4h). Figures 4i and 4j show the temperature dependence of spin-spin correlation function scans at $E = 2.5 \pm 0.5$ meV along the $[H, 0,0]$ and $[1,0, L]$ directions, respectively. Similar scans at $E = 11.5 \pm 0.5$ meV are shown in Figs. 4m and 4n. From the temperature dependent dynamic spin-spin correlation lengths along the in-plane and *c*-axis directions (Figs. 4k and 4o) and magnetic scattering at the $\Gamma$ and $Z$ points for different energies (Figs. 4l and 4p), we conclude that low-energy spin fluctuations have a 2D to 3D crossover approximately below 400 K ($T_S$), consistent with the establishment of the spin nematic phase (Figs. 3n and 3o).

To prove that spin space anisotropy arises from Yb-induced SOC and determine whether the heavier Yb in $YbMnBi_2$ plays a role in the spin nematic phase, we carried out polarized inelastic neutron scattering measurements on $CaMnBi_2$. Figures 5a-5h show the energy dependence of $\sigma^{SF}_{x,y,z}$ for $CaMnBi_2$ measured at $\boldsymbol{Q}_1$ and $\boldsymbol{Q}_3$ across $T_N \approx 270$ K. Below $T_N$ at 230 K, we find a ~6 meV spin gap, $M_c = 0$, and $M_a = M_b > 0$ (Fig. 5i), similar to gapped transverse spin waves in $YbMnBi_2$ (Fig. 3i). At $T = T_N = 270$ K, we have $M_c > M_a = M_b$ below ~4 meV and $M_c = M_a = M_b$ for energies above 4 meV (Fig. 5j), clearly different from $M_c \approx M_a > M_b$ below 4 meV in $YbMnBi_2$ (Fig. 3j) implying no *ab*-plane spin space anisotropy in $CaMnBi_2$. On warming to temperatures above $T_N$, we find $M_c = M_a = M_b$ at all energies investigated, consistent with isotropic paramagnetic scattering (Figs. 5k, 5l). Figures 5m and 5n show the temperature dependence of $M_a, M_b, M_c$ and $M_a - M_b$ at $E \approx 2$ meV, respectively. As temperature decreases towards $T_N$, $M_c$ peaks at $T_N$ as expected due to c-axis moment long-range order, but in-plane magnetic anisotropy $M_a - M_b$ shows no feature across $T_N$, thus ruling out the existence of a spin nematic phase in $CaMnBi_2$. Figures 5o and 5p show similar results at $E \approx 9$ meV, revealing isotropic paramagnetic scattering above $T_N$ and pure transverse spin waves ($M_c = 0$, and $M_a = M_b > 0$) below $T_N$. Since the major difference between $YbMnBi_2$ and $CaMnBi_2$ is the presence of heavy Yb in $YbMnBi_2$, we conclude that the observed spontaneous spin nematic phase and in-plane spin space anisotropy in the former arise from Yb-induced SOC.

The observation of *ab*-plane spin space anisotropy in $YbMnBi_2$ is surprising because the anisotropy seemingly violates the tetragonal symmetry, which should be present according to the experiment. However, it is noteworthy that because the dynamic spin nematic below 400 K is locked to the $\boldsymbol{Q}$ vectors, it does not break tetragonal symmetry globally. Note here that the $C_4$ operations rotate both $\boldsymbol{Q}$ vectors and spins simultaneously in the presence of the spin-orbit interactions. Figures 6a-6c summarize the energy evolution of the magnetic excitation anisotropy as a function of increasing temperature across $T_N$ for $YbMnBi_2$. Figures 6e-6g show the temperature evolution of the $E$ and $\boldsymbol{Q}$ dependent magnetic excitation anisotropy in spin space. The dynamic nematic spin excitations between $\boldsymbol{Q}$ vectors connected by the tetragonal $C_4$ rotational symmetry (Fig. 6f). Also, this is a single phase that does not form different domains, as the spin is in a fluctuating regime.

To confirm the crystal structure of $A\mathrm{MnBi_2}$, we performed single-crystal neutron diffraction to search for potential lattice symmetry breaking. At all temperatures (100 K, 300 K, and 450 K for $\mathrm{YbMnBi_2}$; 100 K and 300 K for $\mathrm{CaMnBi_2}$), the nuclear structure can be refined using the $P4/nmm$ space group, which does not break the inversion symmetry. Although we observed weak $(H, K, 0)$ reflections with $H + K = odd$, which are forbidden by the $P4/nmm$ symmetry (57), the structural component of the scattering is extremely weak. It has no temperature dependence below 300 K (see appendix). We therefore conclude that they arise from small defects in the crystal and are not associated with the $C_4$ in-plane tetragonal lattice symmetry (57). In addition, our refinements show no evidence of Bi deficiency, thereby ruling out the sizable DM interactions as the origin of a possible canted AFM state (43). Similar results are obtained for $\mathrm{CaMnBi_2}$ (see appendix), consistent with recent work (47).

## B. Discussion

We have discovered a dynamic spin nematic phase that spontaneously appears in $\mathrm{YbMnBi_2}$ at a temperature below ~400 K ($T_S \gg T_N$) without breaking the $T$-symmetry, the $C_4$ rotational and inversion symmetries of the underlying lattice (Figs. 6a-6c and 6e-6g). While the in-plane magnetic field-induced AHE in $\mathrm{YbMnBi_2}$ also becomes visible below ~400 K, the nematic phase alone cannot explain the contrasting AHE behavior in $\mathrm{YbMnBi_2}$ and $\mathrm{CaMnBi_2}$ (Fig. 1f). While AHE in $\mathrm{KV_3Sb_5}$ (58) and $\mathrm{CsV_3Sb_5}$ (59) falls within the extrinsic, skew scattering regime (23), AHE in $\mathrm{YbMnBi_2}$ is in the middle of the intrinsic regime with a similar magnitude as the cubic MnSi without inversion symmetry (skyrmions) (60) (Fig. 1g).

Although a minimum spin canting angle of 2º from the *c*-axis aligned moment in Mn, regardless of the microscopic origin, can potentially drive $YbMnBi_2$ into a type-II Weyl state (40-43), our polarized neutron scattering experiments put an upper bound of 0.3º for the possible canting angle, essentially ruling out the mechanisms associated with spin canting and associated ferromagnetism. Since a separate polarized neutron diffraction (47) and our polarized inelastic neutron scattering work find no evidence of spin canting and spin space anisotropy in spin excitations of $CaMnBi_2$, respectively, an exceedingly small but finite AHE observed in the system (Fig. 1f) must have a different origin from that of $YbMnBi_2$, most likely from the *T*-symmetry-breaking by the applied magnetic field commonly seen for many materials (23).

A key difference between $YbMnBi_2$ and $CaMnBi_2$ lies in the *A*-site ions. While Ca has the nonmagnetic 2+ state, our XPS measurements reveal that more than 50% of Yb in $YbMnBi_2$ adopt the 3+ valence state and carry magnetic moments (Fig. 6d). To directly establish the bulk $Yb^{3+}$ CEF scheme, we performed high-energy inelastic neutron scattering measurements on $YbMnBi_2$ using the ARCS spectrometer at ORNL with incident energies $E_i = 200$, 250, and 300 meV (61). The single crystal $YbMnBi_2$ sample used in polarized neutron scattering experiments was cooled to 10 K. Figures 7a–7c show broad $|\boldsymbol{Q}|$-energy maps in which several nondispersive features are visible at high energy transfers. Constant-$|\boldsymbol{Q}|$ cuts, shown in Figs. 7e-7g, reveal three excitations centered at approximately $\Delta_1 = 92.2$, $\Delta_2 = 163.7$, and $\Delta_3 = 180.2$ meV. These modes are essentially nondispersive and their intensities decrease with increasing $|\boldsymbol{Q}|$, as expected for magnetic scattering from localized $4f$ moments following the $Yb^{3+}$ magnetic form factor. The energies of these modes are also well above the known spin waves and phonon band top. For $Yb^{3+}$, the $4f^{13}$ configuration gives a Hund's-rule $J = 7/2$ multiplet, which is split by the CEF into four

Kramers doublets. The three observed excitations therefore correspond naturally to transitions from the ground-state doublet to the three excited doublets (Fig. 7d). In contrast, nonmagnetic $Yb^{2+}$ has a closed-shell $4f^{14}$, $J = 0$ configuration, and therefore cannot produce such crystal electric field excitations. These neutron data provide direct spectroscopic evidence for the presence of $Yb^{3+}$ ions in $YbMnBi_2$.

A CEF calculation constrained by the measured excitation energies and relative integrated intensities yields a ground-state Kramers doublet dominated by the $|J_z| = 1/2$ components, with calculated $g$ factors $(g_a, g_b, g_c) \approx (4.44, 4.44, 0.88)$ and dipolar moments $(\vec{m}_a, \vec{m}_b, \vec{m}_c) = (2.22, 2.22, 0.440)\mu_B$ for the averaged dataset within our probed temperature range. Thus, the $Yb^{3+}$ CEF ground state strongly suppresses the *c* axis dipolar moment relative to the free-ion value, while leaving field-polarizable in-plane magnetic degrees of freedom that can couple to Mn spin correlations. This provides a microscopic basis for the much larger field-induced SSC, AHE, and ANE in $YbMnBi_2$ than in $CaMnBi_2$.

The physical picture of these phenomena in $A$MnBi$_2$ is as follows. The nematic order and Néel orders emerge due to the spontaneous symmetry breaking of the magnetic moments of Mn in spin space and real space, respectively. Considering that half of the Yb atoms have a magnetic moment while Ca atoms do not, the magnetic Yb moments provide the field-polarizable degrees of freedom that enable the enhanced SSC-driven AHE and ANE. Importantly, $Yb^{3+}$ moments play a special role - it forms a triangle with Mn moments and induces the SSC $\chi_{ijk} = \boldsymbol{S}_i \cdot (\boldsymbol{S}_j \times \boldsymbol{S}_k)$ (Fig. 6h), which contributes to the anomalous Hall, anomalous Nernst, and thermal Hall effects by breaking *T*-symmetry (62-65). Thus, only $YbMnBi_2$ can exhibit SSC-origin AHE and ANE. The $Yb^{3+}$ acts as a paramagnetic "spin-catalyst" under an in-plane magnetic field $\boldsymbol{B}$, inducing SSC at all temperatures but dramatically enhancing the SSC when the nematic order emerges because of the SSC-nematic order coupling through $\boldsymbol{B}$ (Fig. 6n).

Such a picture is justified by establishing the Landau theory of order parameters. Let us first consider the system's symmetry and structure. The space group of $A$MnBi$_2$ is $P4/nmm$ (No. 129). The Wyckoff position of Mn ions is $2a$, that of the Yb or Ca ions is $2c$, and that of the Bi ions is $4f$. The unit cell is depicted in Fig. 6h. As two Mn ions and an $A$ ion form a triangle, there are a total of two triangles in the unit cell. Considering the whole structure, each Mn ion is surrounded by eight triangles, as shown in Fig. 6i. We note that the SSC is a pseudo-scalar quantity. Still, one can define the direction of SSC as the direction of the effective magnetic field, *i.e.*, normal to the triangle plane. Therefore, only two directions per triangle can exist. When the SSC configuration does not cancel, the AHE can be induced by the remaining SSC.

Based on Figs. 6h, 6i and experiments, the order parameters are set to be the nematic order $\psi_- = \langle S_{1x}S_{2x} - S_{1y}S_{2y}\rangle = M_a - M_b$, Néel order $N = S_{1z} - S_{2z}$ and the SSC configuration $\boldsymbol{\chi_S} = \left(\chi_{S,x}, \chi_{S,y}, \chi_{S,z}\right)$. We present the SSC configuration in Fig. 6j-6l. Here, $\psi_-$ and $N$ are the main order parameters that appear from the spontaneous symmetry breaking, while $\boldsymbol{\chi_S}$ is the secondary order parameter that is induced by the external magnetic field. We also consider the external magnetic field $\boldsymbol{B} = (B_x, B_y, B_z)$. By investigating the irreducible representations (IRREPs) of the point group $4/mmm$ little group at $\Gamma$ point (66), we find that $\psi_-$ is in $\Gamma_2^+$, $N$ is in $\Gamma_3^+$, $\left(\chi_{S,x}, \chi_{S,y}\right)$ and $(B_x, B_y)$ are in $\Gamma_5^+$, and $\chi_{S,z}$ and $B_z$ are in $\Gamma_3^+$.

Considering both crystalline and time-reversal symmetries, the permitted coupling terms are the following. Up to second order, $\boldsymbol{B} \cdot \boldsymbol{\chi}_S = B_x \chi_{S,x} + B_y \chi_{S,y} + B_z \chi_{S,z}$ are available. These couplings are trivial because the physical meaning is just that the magnetic field induces the SSC. Up to the third order of order parameters, $\psi_-(B_x^2 - B_y^2)$, $\psi_-\left(B_x \chi_{S,x} - B_y \chi_{S,y}\right)$ are available. The first coupling is trivial, as $\psi_-$ is the difference of fluctuations in $x$ and $y$ directions. However, the last coupling term is at the heart of this work; the nematic order couples to the in-plane SSC through the magnetic field.

Here, please note that we consider coupling within each unit cell, so each order parameter depends on the unit cell position. This makes the couplings depending on the position, i.e., $\psi_-(\boldsymbol{r})\left(B_x \chi_{S,x}(\boldsymbol{r}) - B_y \chi_{S,y}(\boldsymbol{r})\right)$. Performing the Fourier transform to the momentum with inversion symmetry, one could notice that the coupling $\psi_-(\boldsymbol{Q})\left(B_x \chi_{S,x}(\boldsymbol{Q}) - B_y \chi_{S,y}(\boldsymbol{Q})\right)$.

Given the classification, the established Landau free energy is written as follows.

$$F = \int d\boldsymbol{Q}\, F_0(\boldsymbol{Q}) + F_1(\boldsymbol{Q}) + F_2(\boldsymbol{Q}), \quad (1)$$

where

$$F_0(\boldsymbol{Q}) = \alpha_2(\boldsymbol{Q}, T)|\psi_-(\boldsymbol{Q})|^2 + \alpha_4|\psi_-(\boldsymbol{Q})|^4 + \beta_2(T)|N(\boldsymbol{Q})|^2 + \beta_4|N(\boldsymbol{Q})|^4, \quad (2)$$

is the free energy for the main order parameters, while

$$F_1 = \eta_1|\chi_{S,x}(\boldsymbol{Q})|^2 + \eta_2\left|\chi_{S,y}(\boldsymbol{Q})\right|^2, \quad (3)$$

is the free energy for the secondary order parameters, and

$$F_2 = g_1\left(B_x\chi_{S,x}(\boldsymbol{Q}) + B_y\chi_{S,y}(\boldsymbol{Q})\right)\delta(\boldsymbol{Q}) + g_2\left(B_x\chi_{S,x}(\boldsymbol{Q}) - B_y\chi_{S,y}(\boldsymbol{Q})\right)\psi_{-}(\boldsymbol{Q}), \tag{4}$$

is the free energy for the couplings between order parameters up to the third order. We consider only the in-plane magnetic field. Here, $\alpha_2(\boldsymbol{Q},T) = \alpha(|\boldsymbol{Q}|)(T - T_S)$ and $\beta_2(T) = T - T_N$ are the temperature differences, $\alpha_4$, $\beta_4$, $\eta_1$, $\eta_2$ are the positive constants, and $g_i$ with $i = 1{,}2$ are arbitrary constants. $g_1$ appears only for $\boldsymbol{Q} = 0$. In absence of magnetic field, when $\alpha(|\boldsymbol{Q}| = 0) = 0$ and $\alpha(|\boldsymbol{Q}| \neq 0) < 0$, there could be finite $\psi_{-}(\boldsymbol{Q})$ for finite $\boldsymbol{Q}$ without breaking $C_4$ symmetry. Based on the observation, additionally, the relation of $T_S$ and $T_N$ is assumed to be $T_S > T_N$. A typical change in order parameters with temperature at a fixed nonzero $\boldsymbol{Q}$ is shown in Figs. 6p and 6q, by fixing $T_S = 1$, $T_N = 0.5$, $\alpha(|Q| \neq 0) = \alpha_4 = \beta_4 = \eta_1 = \eta_2 = 1$, $g_2 = -0.1$, and $\boldsymbol{B} = (1{,}0{,}0)$.

Based on our Landau theory described above, the possible scenarios to explain the experimental results are as follows. Regardless of the temperature, the SSC appears in the in-plane magnetic field. By cooling the system, the spin nematic order emerges first, followed by the Néel order at a lower temperature. When the nematic order emerges, the coupling of in-plane SSC, nematic order, and magnetic field strengthens the in-plane SSC, as shown in Fig. 6n. As a result, the AHE rises in a similar temperature range with the nematic order. On cooling to a temperature well below $T_N$, the gradual opening of a spin gap suppresses the nematic order and therefore the AHE. Because the magnetic structure in $A$MnBi$_2$ is collinear (Fig. 2), static magnetic order cannot induce SSC and AHE.

Accordingly, in YbMnBi$_2$, where magnetic $Yb^{3+}$ ions coexist with the Mn spin system, the field-induced SSC can appear together with the nematic and Néel orders. The associated AHE and ANE are therefore strongly enhanced in $YbMnBi_2$. In contrast, $CaMnBi_2$ contains nonmagnetic $Ca^{2+}$ ions and shows neither a spin nematic phase nor a comparable anomalous transport response. Our results therefore provide compelling evidence for spin-nematic-phase-enhanced SSC and AHE/ANE (40, 41) in the paramagnetic state of a compensated collinear antiferromagnet, different from all other known ways of producing AHE (20, 21, 24, 25, 67-72).

## III. Conclusions

The microscopic origin of the spin nematic phase could be attributed to the presence of four-spin exchange interactions (10) among Mn moments and their interaction with Yb. In $A$MnBi$_2$ systems, the Mn atoms form a square lattice structure. It is well established that in spin models on a square lattice, such as cuprates and iridates, four-spin (ring) exchange interactions naturally arise and can stabilize a spin nematic phase (10, 11). Since the four-spin exchange can be interpreted as an effective quadrupole–quadrupole interaction, spontaneous symmetry breaking occurs when the quadrupole moments are ordered. Because of the presence of heavy Yb, the four-spin exchange interactions are dramatically enhanced in $YbMnBi_2$, giving rise to the observed in-plane spin space anisotropy and spin nematic phase above $T_N$. Although similar four-spin exchange interactions should also be present in $CaMnBi_2$, they are not strong enough to induce in-plane spin space anisotropy. It is important to note that the spin nematic phase is, by definition, a form of quadrupolar order. As we theoretically show that the SSC induced by the combination of an external magnetic field and spin nematic order arises from fifth-order spin couplings, our experiments demonstrate a link between higher-order composite order parameters. The physical consequences of the link between higher-order terms indicate a new direction for research and offer a new avenue towards room-temperature fast spintronics without magnetic order.

**Acknowledgements**

We thank Douglas L Abernathy for his help in crystal electric field measurements of $YbMnBi_2$ on ARCS, SNS, ORNL and appreciate Ms. Mari Ishida for designing Fig. 5h-5n. The single crystal growth and neutron scattering work at Rice is supported US DOE BES DE-SC0012311 and DE-SC0026179 (P.D.). Part of the materials characterization work at Rice is supported by the Robert A. Welch Foundation under Grant No. C-1839 (P.D.). This work was done in part using resources of the Shared Equipment Authority at Rice University (https://research.rice.edu/sea/). We thank Dr. Bo Chen for help with XPS measurements. T.O. was supported by Basic Science Research Program and G-LAMP through the National Research Foundation of Korea (NRF) funded by the Ministry of Education (RS-2021-NR060140, RS-2025-16065011, RS-2025-25441317). N.N. were supported by JSPS KAKENHI Grant Numbers 24H00197 and 24H02231, and the RIKEN TRIP initiative. Y.P. acknowledges the financial support from Scientific Research Innovation Capability Support Project for Young Faculty (SRICSPYF-ZY2025076), National Key R&D Program of China (No. 2025YFF0524500) and the National Natural Science Foundation of China (Grant No. 52401263). This research used resources at the High Flux Isotope Reactor and Spallation Neutron Source, DOE Office of Science User Facilities operated by the Oak Ridge National Laboratory. X.W. acknowledges the partial support by the Laboratory Directed Research and Development (LDRD) of the Neutron Sciences Directorate of ORNL. ORNL is managed by UT-Battelle, LLC, under contract DE-AC05-00OR22725 with the U.S. Department of Energy (DOE). The beam time was allocated to DEMAND on proposal number IPTS- 33540.1; to PTAX on proposal numbers ITPS-32310.1, 33145.1, 33994.1, 34110.1, 34791.1, and 37308.1; to TOPAZ on proposal numbers IPTS-34726.1 and IPTS-34727.1; to HYPSEC on proposal numbers ITPS-34121.1, to ARCS on proposal numbers IPTS-38568.

**Author Contributions**

P.D. and Y.P. conceived the project. Y.X., S.X., Y.P., T.O., and T.Z. made equal contributions to the project. S.X., Y.P., C.F., Z.W., Y.W., S.W., X.Z., made samples. Neutron scattering experiments were carried out and analyzed by Y.X., S.X., M.M., T.Z., A.M., X.W., S.C., F.Y., Y.H., H.C., B.L.W., M.K.G., and P.D.. AHE measurements were performed by S.X., Z.L., Y.P., and C.F.. Theoretical analysis was supervised by N.N., and performed by T.O. and N.N.. The entire project was supervised by P.D. XPS measurements were performed by S.X., S.P., F.L., and X.Z.. The manuscript is written by P.D., Y.X., S.X., T.O., and N.N.. All authors made comments.

Data availability—The data that support the findings of this article are not publicly available upon publication because it is not technically feasible and/or the cost of preparing, depositing, and hosting the data would be prohibitive within the terms of this research project. The data are available from the authors upon reasonable request.

**Appendix:**

**Sample synthesis and composition characterizations**

$CaMnBi_2$ and $YbMnBi_2$ single crystals were grown by a Bi self-flux method, following the procedures described in a previous report (40). Single Crystals with typical dimensions of $4 \times 4 \times 2$ mm$^3$ were obtained. Extended Data Fig. 1 shows Laue X-ray backscattering images of $CaMnBi_2$ and $YbMnBi_2$, highlighting the (0,0,1) surface orientation.

**Transport measurements**

Resistivity and Hall resistivity were measured using a standard five-probe method. Part of the work in $YbMnBi_2$ at Chongqing University was performed using the electrical transport option of the physical property measurement system (PPMS9, Quantum Design). The other work in $CaMnBi_2$ at Rice University was conducted using Lock-in amplifiers (SR830) and with appropriate preamplifiers. The sample was manually polished into a bar shape with typical dimensions of $2.0 \times 0.5 \times 0.2\ mm^3$. The magnetic field is always along the in-plane $a$- (or $b$-) axis. Therefore, the current flows either parallel to the $a$- (or $b$-) axis to obtain $\rho_{bb}$ and $\rho_{cb}$, or parallel to $c$-axis to obtain $\rho_{cc}$ . All field data was symmetrized and antisymmetrized to correct for contact misalignments. The anomalous Hall resistivity was extracted by subtracting the linear background at higher fields, as presented in Extended Data Figs. 2 and 3.

**Polarized neutron scattering experiments**

Neutron LPA experiments on $A$MnBi$_2$ ($A$ = Ca,Yb) were carried out on the polarized triple axis spectrometer PTAX (HB-1) at the High Flux Isotope Reactor (HFIR), Oak Ridge National Laboratory (ORNL). The neutron beam was polarized and analyzed with a Heusler monochromator and a Heusler analyzer with the final energy fixed at $E_f = 13.5$ meV. Pyrolytic graphite filters were used to reduce higher harmonic contamination. The flipping ratio of the polarized neutron beam, defined as $R = NSF_N/SF_N$ (where $NSF_N$ and $SF_N$ are measured NSF and SF scattering cross sections on a nuclear Bragg peak) (2, 8, 48), was about 12 measured on nuclear Bragg peak $(0,0,3)$. The guide field used for neutron polarization is about 15 Gauss, which corresponds to a Zeeman energy $E = g\mu_B B \approx 2 \times 10^{-4}$ meV. This energy scale is negligible compared with the meV spin-fluctuation energies and thermal energies at our measurement temperatures. Therefore, the guide field cannot account for the observed in-plane spin anisotropy.

The momentum transfer $\boldsymbol{Q}$ in reciprocal space is defined as $\boldsymbol{Q} = H\boldsymbol{a}^* + K\boldsymbol{b}^* + L\boldsymbol{c}^*$ with $\boldsymbol{a}^* = (\frac{2\pi}{a})\hat{\boldsymbol{a}}$, $\boldsymbol{b}^* = (\frac{2\pi}{b})\hat{\boldsymbol{b}}$, $\boldsymbol{c}^* = (\frac{2\pi}{c})\hat{\boldsymbol{c}}$, where $a = b = 4.5033$ Å and $c = 10.843$ Å are lattice constants at room temperature, and $H, K, L$ are Miller indices. For the polarized neutron diffraction experiments, where the instrumental energy resolution is about $\Delta E \approx 0 \pm 0.75$ meV in full width half maximum (FWHM), we used one piece of single crystal for each compound: a ~50 mg crystal of $CaMnBi_2$ and a ~250 mg crystal of $YbMnBi_2$, respectively. Both samples are mounted in the $[H, 0, L]$ scattering plane. For the polarized inelastic neutron scattering measurement on $YbMnBi_2$, one single crystal weighing 3.4 grams was used and energy resolution is about $\sim$1.5 meV at elastic line. A guide magnetic field of about 15 Gauss is applied on the sample, and the magnetic guide field direction is always along the momentum transfer $\boldsymbol{Q}$ direction. For polarized inelastic neutron scattering experiments on $CaMnBi_2$, we co-aligned about 20 pieces of single crystals with a total mass of 3.6 grams on an aluminum plate and have the HB-1 spectrometer setup identical to that of the $YbMnBi_2$. The energy resolution is about $\sim$1.5 meV at elastic line. A guide magnetic field of about 15 Gauss is applied on the sample using Helmholtz coils, and the magnetic guide field direction is along the momentum transfer $\boldsymbol{Q}$ direction ($x$), perpendicular to $\boldsymbol{Q}$ in the scattering plane ($y$), or perpendicular to the scattering plane ($z$).

**Magnetic structure factors at $\boldsymbol{Q} = (1,0,0)$, $\boldsymbol{Q} = (2,0,0)$, and $\boldsymbol{Q} = (0,0,1)$**

For magnetic scattering at $\boldsymbol{Q} = (1,0,0)$, the measured neutron cross sections are proportional to the square of the static magnetic structure factor,

$$\boldsymbol{F}_M(\boldsymbol{G}_M) = \sum_j f(\boldsymbol{Q})\boldsymbol{S}_{\perp j} e^{\boldsymbol{G}_M \cdot \boldsymbol{d}_j}$$

where $f(\boldsymbol{Q})$ is the magnetic form factor, and $\boldsymbol{S}_{\perp j}$ is the component of the spin at site $\boldsymbol{d}_j$ perpendicular to $\boldsymbol{Q}$. The magnetic unit cell is composed of two Mn ions at positions $\boldsymbol{d}_1$= (0,0,0) and $\boldsymbol{d}_2$ = (0.5,0.5,0) with spin amplitudes $(S_{a1}, S_{b1}, S_{c1})$ and $(S_{a2}, S_{b2}, S_{c2})$, respectively. For $\boldsymbol{Q} = (H, 0,0)$, the squared magnitude of the magnetic structure factor becomes

$$|\boldsymbol{F}_M|^2 = |f(\boldsymbol{Q})|^2(S_{b1}^2 + S_{b2}^2 + 2S_{b1}S_{b2}\cos H\pi) + |f(\boldsymbol{Q})|^2(S_{c1}^2 + S_{c2}^2 + 2S_{c1}S_{c2}\cos H\pi)$$

In the case of LPA neutron measurements, the magnetic responses $M_y$ and $M_z$ (along the $y$ and $z$ directions, respectively) are proportional to the terms involving $S_c$ and $S_b$. Specifically,

$$M_y \sim S_{c1}^2 + S_{c2}^2 + 2S_{c1}S_{c2}\cos H\pi\,, M_z \sim S_{b1}^2 + S_{b2}^2 + 2S_{b1}S_{b2}\cos H\pi.$$

At $\boldsymbol{Q} = (1,0,0)$, these expressions simplify to $M_y \sim M_c \sim (S_{c1} - S_{c2})^2$ and $M_z \sim M_b \sim (S_{b1} - S_{b2})^2$, which measures the AFM components along *c* and *b* axis, respectively. In contrast, at $\boldsymbol{Q} = (2,0,0)$, $M_y \sim M_c \sim (S_{c1} + S_{c2})^2$ and $M_z \sim M_b \sim (S_{b1} + S_{b2})^2$, which measure the net ferromagnetic components along *c* and *b* axis, respectively. These relations thus enable us to distinguish AFM and ferrimagnetic contributions in each spin direction.

If any in-plane magnetic component (regardless of its in-plane direction) is present, the squared magnitude of the magnetic structure factor at $\boldsymbol{Q} = (0,0,1)$ is described as

$$|\boldsymbol{F}_M|^2 = |f(\boldsymbol{Q})|^2(S_{a1} + S_{a2})^2 + |f(\boldsymbol{Q})|^2(S_{b1} + S_{b2})^2$$

Then the magnetic responses $M_y$ and $M_z$,

$$M_y \sim (S_{b1} + S_{b2})^2, M_z \sim (S_{a1} + S_{a2})^2$$

**Estimating the upper limit of the in-plane Mn magnetic moments**

To estimate the upper limit of the in-plane magnetic moment, we compare the integrated intensities at two Bragg reflections: $\boldsymbol{Q} = (0,0,1)$, which is sensitive to the in-plane magnetic moments, and $\boldsymbol{Q} = (1,0,0)$, which primarily probes the *c*-axis magnetic moment. The measured integrated intensity at a given $\boldsymbol{Q}$ is related to the squared magnitude of the magnetic structure factor through:

$$\mathrm{Int}(\boldsymbol{Q}) \propto |\boldsymbol{F}_M(\boldsymbol{Q})|^2 / \sin 2\theta_{\boldsymbol{Q}}$$

where $\theta_{\boldsymbol{Q}}$ is the Bragg angle of the reflection. Then the ratio of squared magnetic structure factors for the two reflections simplifies to

$$\frac{|\boldsymbol{F}_M(0,0,1)|^2}{|\boldsymbol{F}_M(1,0,0)|^2} = \frac{\mathrm{Int}_{(0,0,1)} \sin 2\theta_{(0,0,1)}}{\mathrm{Int}_{(1,0,0)} \sin 2\theta_{(1,0,0)}}$$

Given that no significant magnetic signal is observed at $\boldsymbol{Q} = (0,0,1)$, we take the measured background intensity—$M_y(0,0,1) = \sigma_x^{\mathrm{SF}} - \sigma_y^{\mathrm{SF}}$ and $M_z(0,0,1) = \sigma_x^{\mathrm{SF}} - \sigma_z^{\mathrm{SF}}$, as shown in Fig. 2d— as an upper bound. This is then compared to the magnetic intensity at $\boldsymbol{Q} = (1,0,0)$, where $M_y(1,0,0) = \sigma_x^{\mathrm{SF}} - \sigma_y^{\mathrm{SF}}$ (see Fig. 2e), which reflects the *c*-axis moment. Assuming that the moment size is uniform across different Mn sites, and using the experimentally determined *c*-axis moment of 3.3 $\mu_B$ at 50 K, along with the measured integrated intensities: $M_y(0,0,1) = -4.02 \times 10^{-8} \pm 7.11 \times 10^{-6}$, $M_z(0,0,1) = 3.02 \times 10^{-5} \pm 7.10 \times 10^{-6}$, and $M_y(1,0,0) = 1.17 \times 10^5 \pm 8.89 \times 10^2$, we obtain conservative upper limits of 0.009 and 0.019 $\mu_B$ for the moment components along the *a* and *b* axes respectively. Based on these values, the corresponding upper limits for the canting angles of the magnetic moment toward the *a* and *b* axes are approximately 0.15° and 0.33°, respectively. These results are consistent with polarized neutron diffraction experiments on $CaMnBi_2$ (47).

**Determination of $M_a$, $M_b$, and $M_c$ in spin excitations of $YbMnBi_2$** and $CaMnBi_2$

Magnetic neutron scattering probes only the magnetic moment component perpendicular to $\boldsymbol{Q}$. The measured cross sections can be expressed in terms of magnetic response along the $y$ ($M_y$) and $z$ ($M_z$) directions. However, due to imperfect neutron polarization, there is a leakage between $SF$ and $NSF$ channels. This leakage can be quantified using the flipping ratio $R = NSF_N/SF_N \approx 12$, where $NSF_N$ and $SF_N$ of $(0,0,3)$ are nuclear Bragg peak intensities measured in $NSF$ and $SF$ channels, respectively. The measured cross sections can then be written(8)

$$\sigma_x^{\mathrm{SF}} = \frac{R}{R+1}M_y + \frac{R}{R+1}M_z + \frac{1}{R+1}NC + \frac{2R+1}{3(R+1)}NSI + BG$$

$$\sigma_y^{\mathrm{SF}} = \frac{1}{R+1}M_y + \frac{R}{R+1}M_z + \frac{1}{R+1}NC + \frac{2R+1}{3(R+1)}NSI + BG$$

$$\sigma_z^{\mathrm{SF}} = \frac{R}{R+1}M_y + \frac{1}{R+1}M_z + \frac{1}{R+1}NC + \frac{2R+1}{3(R+1)}NSI + BG$$

where $NC$ is the nuclear coherent scattering, $BG$ is the instrumental background scattering, which is assumed to be the same for all cross sections at a given wavevector and energy transfer. *NSI* is the nuclear spin incoherent scattering, which is negligible compared to $NC$ in these compounds. For an arbitrary $\boldsymbol{Q}$ within $[H, 0, L]$ scattering plane (Fig. 3m), one can probe magnetic response within the *y*-*z* plane, giving $M_y = M_a\cos^2\theta + M_c\sin^2\theta$ and $M_z = M_b$, where $\theta$ is the angle between $\boldsymbol{Q}$ and $[0,0,L]$. By probing two or more equivalent AFM wave vectors with different angle $\theta$, we can conclusively determine $M_a$, $M_b$, and $M_c$ (Fig. 3i-3l), via

$$\sigma_x^{\mathrm{SF}}(\boldsymbol{Q}_1) = f^2(\boldsymbol{Q}_1)\left(\cos^2\theta_{\mathbf{Q}_1}\frac{R}{R+1}M_a + \frac{R}{R+1}M_b + \sin^2\theta_{\mathbf{Q}_1}\frac{R}{R+1}M_c\right) + BG(\boldsymbol{Q}_1)$$

$$\sigma_y^{\mathrm{SF}}(\boldsymbol{Q}_1) = f^2(\boldsymbol{Q}_1)\left(\cos^2\theta_{\mathbf{Q}_1}\frac{1}{R+1}M_a + \frac{R}{R+1}M_b + \sin^2\theta_{\mathbf{Q}_1}\frac{1}{R+1}M_c\right) + BG(\boldsymbol{Q}_1)$$

$$\sigma_z^{\mathrm{SF}}(\boldsymbol{Q}_1) = f^2(\boldsymbol{Q}_1)\left(\cos^2\theta_{\mathbf{Q}_1}\frac{R}{R+1}M_a + \frac{1}{R+1}M_b + \sin^2\theta_{\mathbf{Q}_1}\frac{R}{R+1}M_c\right) + BG(\boldsymbol{Q}_1)$$

$$\sigma_x^{\mathrm{SF}}(\boldsymbol{Q}_3) = rf^2(\boldsymbol{Q}_2)\left(\cos^2\theta_{\mathbf{Q}_3}\frac{R}{R+1}M_a + \frac{R}{R+1}M_b + \sin^2\theta_{\mathbf{Q}_3}\frac{R}{R+1}M_c\right) + BG(\boldsymbol{Q}_3)$$

$$\sigma_y^{\mathrm{SF}}(\boldsymbol{Q}_3) = rf^2(\boldsymbol{Q}_3)\left(\cos^2\theta_{\mathbf{Q}_3}\frac{1}{R+1}M_a + \frac{R}{R+1}M_b + \sin^2\theta_{\mathbf{Q}_3}\frac{1}{R+1}M_c\right) + BG(\boldsymbol{Q}_3)$$

$$\sigma_z^{\mathrm{SF}}(\boldsymbol{Q}_3) = rf^2(\boldsymbol{Q}_3)\left(\cos^2\theta_{\mathbf{Q}_3}\frac{R}{R+1}M_a + \frac{1}{R+1}M_b + \sin^2\theta_{\mathbf{Q}_3}\frac{R}{R+1}M_c\right) + BG(\boldsymbol{Q}_3)$$

where $R \approx 12$ is the flipping ratio, $f(\boldsymbol{Q})$ is the magnetic form factor of $Mn^{2+}$, $BG(\boldsymbol{Q})$ is the wave vector dependent but polarization independent background, and $\theta(\boldsymbol{Q})$ is the angle between $\boldsymbol{Q}$ and $[0,0,L]$, and *r* is the intensity ratio factor between the two equivalent wave vectors which accounts for the differences in sample illumination volume and the convolution with instrumental resolution (55). The six equations can be used to determine the six unknowns $M_a$, $M_b$, $M_c$, $B(\boldsymbol{Q_1})$, $BG(\boldsymbol{Q_3})$ and *r* shown in Figs. 3i-3l using the raw data shown in Extended Data Fig. 8. Specifically, $M_b$ can be obtained directly from the difference $\sigma_x^{\mathrm{SF}} - \sigma_z^{\mathrm{SF}}$ at $Q_1$ or $Q_2$, whereas $M_a$ or $M_c$ is a linear combination of the measured differences $\sigma_x^{\mathrm{SF}} - \sigma_y^{\mathrm{SF}}$ at $Q_1$ and $Q_2$, weighted with known parameter $\theta$, magnetic form factors, and most importantly, the intensity ration factor, which can be obtained as

$$r = \frac{f^2(Q_1)(\,\sigma_x^{\mathrm{SF}}(Q_2) - \sigma_z^{\mathrm{SF}}(Q_2))}{f^2(Q_2)(\,\sigma_x^{\mathrm{SF}}(Q_1) - \sigma_z^{\mathrm{SF}}(Q_1))}$$

We note that deviations of the magnetic form factor from that of $Mn^{2+}$ are absorbed into *r*, our analysis is independent of the specific values assumed for the magnetic form factor. Therefore, the reliability of $M_a$ or $M_c$ ultimately depends on the robustness of the intensity ratio factor *r*. The reliability of *r* is not satisfied when $\sigma_x^{\mathrm{SF}} \approx \sigma_z^{\mathrm{SF}}$. To avoid this issue, we only fit *r* at positions where $\sigma_x^{\mathrm{SF}}$ and $\sigma_z^{\mathrm{SF}}$ differ significantly, and then extrapolate the fitted values to other points where $\sigma_x^{\mathrm{SF}} \approx \sigma_z^{\mathrm{SF}}$. As shown in Extended Data Fig. 9, the resulting *r* is well described by a constant over the energy and temperature range of our measurements. For *r* in the range between 0.7 and 0.82, the extracted values of $M_a$, $M_b$, $M_c$, and spin fluctuation anisotropy $M_a - M_b$ remain qualitatively robust.

Our polarized inelastic neutron scattering experiments on $CaMnBi_2$ were carried out on the same spectrometer with identical experimental setup to allow a direction comparison between these two systems. Extended Data Fig. 10 shows the energy dependent *r* values from the measurements.

**Unpolarized inelastic neutron scattering**

Inelastic neutron scattering experiments on single-crystal $YbMnBi_2$ were performed on the time-of-flight direct geometry Hybrid Spectrometer (HYSPEC) at the Spallation Neutron Source (SNS), Oak Ridge National Laboratory (ORNL)(73). A crystal weighing approximately 3 grams was aligned in the $[H, 0, L]$ scattering plane and mounted in a bottom-loading, high-temperature closed-cycle refrigerator. The same crystal was previously used for inelastic polarized neutron scattering measurements at HB1, HFIR, ORNL. Measurements were conducted at temperatures of 270 K, 300 K, 350 K, 400 K, 425 K, and 450 K, using incident neutron energies of 7.5 meV and 25 meV. Some of the data are summarized in Fig. 4.

**Neutron single crystal diffraction**

The neutron single crystal diffraction measurements on $YbMnBi_2$ and $CaMnBi_2$ were carried out on the time-of-flight single crystal diffractometer, TOPAZ (74), at SNS, ORNL. High-resolution single-crystal diffraction data with more than 1000 Bragg peaks were collected at 450 K, 300 K, and 100 K for $YbMnBi_2$ and 300 K and 100 K for $CaMnBi_2$. CrystalPlan (75) was employed to optimize sample orientations for data collection, each lasting approximately 1 hour, with 5 C of proton charge for SNS beam power at 1.7 MW. A multiresolution machine learning approach was employed in the 3D reciprocal space in $(H, K, L)$ to obtain raw integrated peak intensities. Data normalization, including neutron time-of-flight spectrum, Lorentz, and detector efficiency corrections, was performed following the procedures reported previously(76). The reduced data were saved in SHELX HKLF2 format, recording the wavelength separately for each reflection, and were not merged. Nuclear and Magnetic structure refinements were performed using the JANA2020 program (77). At 450 K, the nuclear structure of $YbMnBi_2$ was refined in the space group *P4/nmm*, consistent with previous reports (37, 38, 45, 46). At 300 K, weak $(H, K, 0)$ reflections with $H + K = odd$, violating the $n$-glide symmetry of *P4/nmm*, were observed, suggesting a symmetry lowering to the tetragonal *P4mm* or $P{-}4m2$ space group. Neutron LPA measurements on $YbMnBi_2$ at $\boldsymbol{Q} = (1,0,0)$ (Extended Data Fig. 5) suggests a small (about 0.3% intensity of the (2,0,0) nuclear Bragg peak) but discernible structural scattering at 300 K. While the observed nuclear scattering at $(1,0,0)$ violates the systematic absence conditions for *P4/nmm* in the nuclear structure (57), the scattering likely arises from structural defects in the sample and has no temperature dependence below 300 K (Extended Data Fig. 5). Therefore, there is no need to lowering the crystal symmetry to model the magnetic structure. In the refined magnetic space group *P4'/n'm'm*, the AFM ordering also generates additional reflections such as (1,0,0), (1,2,0), and (3,0,0) that are forbidden by the nuclear space group *P4/nmm*, specifically those violating the *n*-glide absence conditions. The magnetic peaks observed in neutron diffraction are caused by the component of the magnetic moment perpendicular to the scattering vector $\boldsymbol{Q}$. The appearance of these reflections is consistent with the projection of the magnetic moment perpendicular to the scattering vector in the *ab*-plane of the tetragonal cell, as observed in the TOPAZ data at 100 K (some also appear above room temperature). Refined lattice parameters and atomic positions are summarized in Extended Data Tables 1-4.

We refined the Bi occupancy at both crystallographic sites using datasets collected at TOPAZ. For the $YbMnBi_2$ sample, refinements at all three temperatures show no evidence of Bi deficiency at the Bi1 site. At 450 K and 300 K, although refinements suggest a small deficiency at the Bi2 site, refinement against the 100 K dataset (in the magnetically ordered state) yields no evidence of Bi deficiency (Extended Table 2). For the $CaMnBi_2$ sample, deficiencies of approximately 2% are observed at the Ca and Mn sites rather than at the Bi sites (Extended Table 3). Moreover, for all datasets collected for both compounds, enabling occupancy refinement of the Bi sites results in negligible improvement in the refinement quality, as indicated by the very small changes in R, wR2, and goodness-of-fit values (Extended Tables 2 and 3). Therefore, we conclude that neutron diffraction provides no evidence for Bi deficiency in these compounds within 0.5% uncertainty of the measurements.

Extended Data Tables 1-4

1. Laue X-ray pattern, Structural refinement, EDX spectroscopy of Yb/$CaMnBi_2$
2. The following table shows the refinement results with the occupancy of Bi enabled for the $YbMnBi_2$ sample.

| Temperature | 100 K | 300 K | 450 K |
| --- | --- | --- | --- |
| Space group (Magnetic Space group) | P4/nmm (No.129) P4'/n'm'm | P4/nmm (No.129) | P4/nmm (No.129) |
| Cell parameters | $a = 4.4923(4)$<br>$b = 4.4923(4)$<br>$c = 10.8388(14)$<br>$\alpha = 90$<br>$\beta = 90$<br>$\gamma = 90$<br>$V = 218.73(4)$ | $a = 4.5033(6)$<br>$b = 4.5033(6)$<br>$c = 10.8430(18)$<br>$\alpha = 90$<br>$\beta = 90$<br>$\gamma = 90$<br>$V = 219.89(6)$ | $a = 4.5087(7)$<br>$b = 4.5087(7)$<br>$c = 10.8740(21)$<br>$\alpha = 90$<br>$\beta = 90$<br>$\gamma = 90$<br>$V = 221.05(7)$ |
| F (000) | 51.528 | 51.528 | 51.528 |
| Index range | $-8 \leq H \leq 8$<br>$-8 \leq K \leq 7$<br>$-21 \leq L \leq 21$ | $-7 \leq H \leq 8$<br>$-8 \leq K \leq 8$<br>$-21 \leq L \leq 21$ | $-8 \leq H \leq 8$<br>$-8 \leq K \leq 7$<br>$-21 \leq L \leq 21$ |
| Reflections collected(N(obs)/N(all)) | 7990/8594 | 6477/6492 | 3303/3315 |
| Final R indices (I > 3*sigma(I)) | R=3.62, wR2=9.09 | R=5.11, wR2=12.29 | R=5.22, wR2=12.44 |
| R indices (all data) | R=4.11, wR2=9.95 | R=5.13, wR2=12.31 | R=5.23, wR2=12.46 |
| Goodness-of-fit | 1.126 | 1.600 | 1.582 |
| Extinction factor | 0.00031 | 0.00023 | 0.000025 |

| Moment size | 3.30273 $\mu_B$ | N/A | N/A |
|---|---|---|---|

100 K:

| | Site | occ | X | y | Z | U11 | U22 | U33 | Ueq |
|---|---|---|---|---|---|---|---|---|---|
| Yb1 | 2c | 1.0 | -0.50 | 1.00 | 0.73190(1) | 0.00918(4) | 0.00917(7) | 0.00924(5) | 0.00920(2) |
| Mn1 | 2a | 1.0 | 0.00 | 1.00 | 0.00 | 0.00801(9) | 0.00801(1) | 0.0103(1) | 0.00878(6) |
| Bi1 | 2b | 1.0 | 0.00 | 1.00 | 0.50 | 0.00732(5) | 0.00732(5) | 0.01042(7) | 0.00836(3) |
| Bi2 | 2c | 1.0 | -0.50 | 1.00 | 0.16528(2) | 0.00701(4) | 0.00701(1) | 0.01010(6) | 0.00804(3) |

300 K:

| | Site | Occ | X | y | Z | U11 | U22 | U33 | Ueq |
|---|---|---|---|---|---|---|---|---|---|
| Yb1 | 2c | 1.0 | 0.25 | 0.25 | 0.73167(2) | 0.1684(7) | 0.01684(7) | 0.01801(9) | 0.01723(4) |
| Mn1 | 2a | 1.0 | 0.75 | 0.25 | 0.00 | 0.0167(2) | 0.0167(2) | 0.0187(3) | 0.0173(1) |
| Bi1 | 2b | 1.0 | 0.75 | 0.25 | 0.50 | 0.01423(8) | 0.01423(8) | 0.0204(1) | 0.01630(6) |
| Bi2 | 2c | 0.974(2) | 0.75 | 0.25 | 0.16536(3) | 0.01321(9) | 0.01321(9) | 0.0180(1) | 0.01479(6) |

450 K:

| | Site | Occ | X | y | Z | U11 | U22 | U33 | Ueq |
|---|---|---|---|---|---|---|---|---|---|
| Yb1 | 2c | 1.0 | 0.25 | 0.25 | 0.73156(3) | 0.212(1) | 0.0212(1) | 0.0220(1) | 0.02150(7) |
| Mn1 | 2a | 1.0 | 0.75 | 0.25 | 0.00 | 0.0209(3) | 0.0209(3) | 0.0234(4) | 0.0217(2) |
| Bi1 | 2b | 1.0 | 0.75 | 0.25 | 0.50 | 0.0177(1) | 0.0177(1) | 0.0253(2) | 0.02024(9) |
| Bi2 | 2c | 0.980(3) | 0.75 | 0.25 | 0.16539(4) | 0.0167(1) | 0.0167(1) | 0.0224(2) | 0.0186(1) |

The following table shows the refinement results with the assumption that the occupancy of all atoms is 100% for the $YbMnBi_2$ sample.

| Temperature | 100 K | 300 K | 450 K |
|---|---|---|---|
| Final R indices (I > 3*sigma(I)) | R=3.62, wR2=9.09 | R=5.14, wR2=12.37 | R=5.23, wR2=12.43 |
| R indices (all data) | R=4.11, wR2=9.95 | R=5.16, wR2=12.39 | R=5.24, wR2=12.45 |
| Goodness-of-fit | 1.126 | 1.609 | 1.581 |

3. Single Crystal Neutron diffraction at TOPAZ ($CaMnBi_2$). The following table shows the refinement results with the occupancy of Ca and Mn enabled for the $CaMnBi_2$ sample.

| | |
|---|---|
| Temperature | 300 K |
| Space group | P4/nmm (No.129) |
| Cell parameters | $a = 4.4920(1)$ |
| | $b = 4.4920(1)$ |
| | $c = 11.0902(4)$ |
| | $\alpha = 90$ |
| | $\beta = 90$ |
| | $\gamma = 90$ |
| | $V = 223.78(1)$ |
| F (000) | 36.068 |
| Index range | $-8 \leq H \leq 4$ |
| | $-7 \leq K \leq 8$ |
| | $-21 \leq L \leq 21$ |
| Reflections collected(N(obs)/N(all)) | 2701/2708 |
| Final R indices (I > 3*sigma(I)) | R=5.38, wR2=12.88 |
| R indices (all data) | R=5.39, wR2=12.88 |
| Goodness-of-fit | 1.727 |
| Extinction factor | N/A |

300 K:

|  | Site | Occ | X | y | Z | U11 | U22 | U33 | Ueq |
|---|---|---|---|---|---|---|---|---|---|
| Ca1 | 2c | 0.983(4) | 0.25 | 0.25 | 0.73036(6) | 0.0178(2) | 0.0178(2) | 0.0179(3) | 0.0178(1) |
| Mn1 | 2a | 0.979(6) | 0.75 | 0.25 | 0.00 | 0.0151(2) | 0.0151(2) | 0.0157(3) | 0.0153(2) |
| Bi1 | 2b | 1.0 | 0.75 | 0.25 | 0.50 | 0.01340(9) | 0.01340(9) | 0.0191(1) | 0.01532(7) |
| Bi2 | 2c | 1.0 | 0.75 | 0.25 | 0.16221(3) | 0.01230(9) | 0.01230(9) | 0.0165(1) | 0.01369(6) |

The following table shows the refinement results with the assumption that the occupancy of all atoms is 100% for the $YbMnBi_2$ sample.

|  |  |
|---|---|
| Temperature | 300 K |
| Final R indices (I > 3*sigma(I)) | R=5.40, wR2=12.96 |
| R indices (all data) | R=5.41, wR2=12.97 |
| Goodness-of-fit | 1.738 |

On one hand, the small deviations in refined occupancies (~2%) do not constitute conclusive evidence for real vacancies within least-squares refinement. The occupancy of a given site is strongly correlated with other parameters, including occupancies of other sites, the overall scale factor, atomic displacement parameters, and absorption and extinction corrections.

On the other hand, comparing these refinement models shows that enabling occupancy refinement does not lead to any significant improvement in refinement quality, as indicated by the R, wR2, and goodness-of-fit values. Therefore, there is no statistically meaningful improvement, and no conclusive evidence for Bi deficiency within the sensitivity of the present single-crystal neutron diffraction measurements. In conclusion, the TOPAZ datasets do not demand Bi deficiency to explain the measured intensities.

**Chemical analysis and $Yb^{3+}$ concentration estimation**

Chemical composition analysis was carried out using energy-dispersive X-ray spectroscopy (EDX) on a FEI Helios NanoLab 660 FIB-SEM system, confirming the homogeneity and stoichiometric ratios of the crystals. XPS using Thermo Fisher Scientific ESCALAB 250Xi, Al Kα was employed to analyze the chemical state of the sample. Argon ion sputtering was performed at an energy of 2 keV, with a total sputtering depth of 60 nm. Each sputtering step removed 6 nm of material with 100 secs, and the process was repeated ten times. XPS data were collected with 30 scans per spectrum, and the energy step size for each scan was 0.1 eV, as shown in Extended Data Fig. 10.

To estimate the amount of $Yb^{3+}$ in $YbMnBi_2$, we first need to assign the $Yb^{2+}$ and $Yb^{3+}$ peaks in Fig. 6d. According to previous work (78), the Yb 4d spectrum of a divalent component exhibits two peaks: a peak near 190 eV shown by green color is assigned to the $4d_{3/2}$ components of $Yb^{2+}$, while a small shoulder located at 180.3 eV corresponding to the $4d_{5/2}$ components of $Yb^{2+}$ can be hardly observed because it is too small (Fig. 6d). The 4d spectrum of $Yb^{3+}$, which is strongly affected by Coulomb and exchange interaction, show multiple peaks as highlighted in blue color of Fig. 6d. The amount of each component can be obtained from calculating the area of the corresponding peaks. This process is conducted by using the Avantage software, which can directly give the corresponding concentration ratio of $Yb^{2+}$ to $Yb^{3+}$.

It is true that compounds with Bi at the cation position (including $YbMnBi_2$) can be oxidized easily. Since we are aware of the oxidation issue, we always cleave the sample for a fresh surface for characterization and transport properties measurements. For the XPS experiments, we first cleave the sample to get a fresh surface, then etching the cleaved surface to further eliminate the oxidization issue. As shown in Extended Data Figure 11b, Bi oxidization peaks are observed on the surface, but none of any oxide peaks are observed in the inside even for an etching depth of only 6 nm. From 6 nm to 60 nm, the peaks are the same. Therefore, we conclude that the "oxidized surface layer" is as thin as less than 6 nm of the freshly cleaved sample, and the existence of $Yb^{3+}$ should be in the bulk with no difference between 6 nm to 60 nm.

Moreover, we note that the main peak of $Yb^{3+}$ is located at ~183 eV, as shown in the Extended Data Fig. 11a. The peak at ~183 eV is much stronger in the inside of the sample (with etching depth larger than 6 nm) than that on the surface (with etching depth of 0 nm), which indicates that the $Yb^{3+}$ is not a result from surface oxidation.

**Inelastic neutron scattering on ARCS to study crystal electric field levels of Yb in $YbMnBi_2$**

To directly establish the valence state and crystal-field scheme of the Yb ions, we performed high-energy inelastic neutron scattering measurements on $YbMnBi_2$ at 10 K using the ARCS spectrometer at the Spallation Neutron Source, ORNL. ARCS is a wide angular-range direct-geometry chopper spectrometer optimized for high neutron flux and large detector coverage, making it well suited for measuring high-energy magnetic excitations and crystal-electric-field transitions (61). All measurements were made with the ARCS-100-1.5 Fermi chopper slit package operating at 600 Hz. This package has a 1.5mm spacing between neutron absorbing slats set to a radius of 0.58m, which provides optimum transmission for 100 meV neutrons when spinning at the maximum rate of 600 Hz. The measurements were carried out with incident energies $E_i = 200, 250, 300$ meV. Three excitations centered at approximately 92.2, 163.7, and 180.2 meV are well above the known spin-wave and phonon energy cutoffs in $YbMnBi_2$, excluding their assignment to conventional collective magnetic or lattice excitations. Importantly, because the first excited crystal electric field doublet lies at ~92.2 meV, corresponding to an energy scale of approximately 1060 K, the $Yb^{3+}$ ions remain confined to the ground-state Kramers doublet over the experimentally relevant temperature range below 450 K. Thus, the low-energy $Yb^{3+}$ degrees of freedom can be represented by an effective pseudospin-1/2 ground-state doublet in the temperature regime relevant to the magnetization and neutron scattering measurements.

Although Mn $3d$ ions can in principle exhibit crystal-field multiplet excitations, these are fundamentally different from rare-earth crystal electric field transitions and typically occur on the scale of hundreds of meV to several eV. In particular, high-spin $Mn^{2+}$ has a $3d^5$, $S = 5/2$, orbital-singlet ${}^6A_g$ ground state and does not generate a sequence of low-energy Kramers-doublet CEF transitions. Therefore, the three nondispersive modes observed at 92.2, 163.7, and 180.2 meV, together with their decreasing intensity with increasing $|\boldsymbol{Q}|$, are naturally assigned to transitions between the four Kramers doublets of the $Yb^{3+}$ $J = 7/2$multiplet rather than to Mn-derived excitations.

To obtain the experimental inputs for the crystal electric field calculation, we performed independent fits to the constant-$|Q|$ cuts from $E_\mathrm{i}$ = 250 and 300 meV datasets, where all three excitations are resolved. The cuts were integrated over $3.5 \leq |Q| < 6.5$ Å$^{-1}$ and fitted using three Gaussian peaks together with a smooth background consisting of a linear term and an exponentially decaying term, as shown in Extended Data Fig. 14. The fitted peak centers are 91.0, 163.4, and 180.2 meV for the $E_\mathrm{i}$ = 250 meV dataset and 93.5, 163.9, and 180.1 meV for the $E_\mathrm{i}$ = 300 meV dataset.

The integrated areas of the three Gaussian components were used to estimate the relative integrated spectral weights of the three crystal electric field transitions. After normalizing the intensity of the second excitation to unity, the relative intensities are 0.11 : 1 : 0.74 for the $E_\mathrm{i}$ = 250 meV dataset and 0.22:1:0.91 for the $E_\mathrm{i}$ = 300 meV dataset. For the CEF analysis, we used the averages of the values independently extracted from the two datasets, yielding excitation energies of 92.2, 163.7, and 180.2 meV and relative intensities of 0.17:1:0.83. The close agreement of the fitted excitation energies, together with the comparable relative spectral weights obtained from the two datasets, supports the use of these averaged values in the CEF analysis.

Using these experimentally determined excitation energies and relative intensities, we modeled the $Yb^{3+}$ CEF scheme. $YbMnBi_2$ crystallizes in the space group P4/nmm (No. 129). with Yb at Wyckoff position 2*c* and site symmetry 4mm ($C_{4v}$).

We modeled the $Yb^{3+}$ CEF scheme using the tetragonal Stevens Hamiltonian

$$H_{CEF} = B_2^0 O_2^0 + B_4^0 O_4^0 + B_4^4 O_4^4 + B_6^0 O_6^0 + B_6^4 O_6^4.$$

We use three sets of the energy splitting and the relative intensity from above: i) $E_i$ = 250 meV, ii) $E_i$ = 300 meV, iii) their average. The fitting reveals the coefficients

$$(1)\ (B_2^0, B_4^0, B_4^4, B_6^0, B_6^4) = (1.07, -0.114, 0.112, 0.00465, 0.00447)\ \text{meV};$$
$$(2)\ (B_2^0, B_4^0, B_4^4, B_6^0, B_6^4) = (1.04, -0.112, 0.131, 0.00472, 0.00779)\ \text{meV};$$
$$(3)\ (B_2^0, B_4^0, B_4^4, B_6^0, B_6^4) = (1.05, -0.113, 0.123, 0.00469, 0.00627)\ \text{meV}.$$

The associated ground state is given by

$$(1)\ 0.990\left|\pm\tfrac{1}{2}\right\rangle - 0.141\left|\mp\tfrac{7}{2}\right\rangle;$$
$$(2)\ 0.982\left|\pm\tfrac{1}{2}\right\rangle - 0.191\left|\mp\tfrac{7}{2}\right\rangle;$$
$$(3)\ 0.986\left|\pm\tfrac{1}{2}\right\rangle - 0.169\left|\mp\tfrac{7}{2}\right\rangle;$$

Thus, the g-factors in each direction are

$$(1)\ (g_a, g_b, g_c) = (4.48, 4.48, 0.961);$$
$$(2)\ (g_a, g_b, g_c) = (4.40, 4.40, 0.808);$$
$$(3)\ (g_a, g_b, g_c) = (4.44, 4.44, 0.881).$$

The dipolar moment is given by

$$(1)\ (\vec{m}_a, \vec{m}_b, \vec{m}_c) = (2.24, 2.24, 0.480)\mu_B;$$
$$(2)\ (\vec{m}_a, \vec{m}_b, \vec{m}_c) = (2.20, 2.20, 0.404)\mu_B;$$
$$(3)\ (\vec{m}_a, \vec{m}_b, \vec{m}_c) = (2.22, 2.22, 0.440)\mu_B.$$

The dipolar moment along the *c* axis is greatly reduced compared to the free-ion $Yb^{3+}$ value.

**Magnetic susceptibility measurements on $YbMnBi_2$**

In Extended Data Figure 12, we present temperature dependence of the magnetization curve. Extended Data Figures 12a and b are the results with the applied magnetic field in the *ab*-plane, measured at Chongqing University, and Extended Data Figure 12c shows the results with the applied magnetic field along the *c* axis, measured at Rice University, respectively. A sharp increase of the magnetization between 200 - 300 K is observed in the field-cooling curves for both *B//ab* and *B//c*. Extended Data Figure 12b presents the effective moment, which is$\sim 0.8 \times 10^{-3}\ \mu_B$ per unit cell.

This small net moment may appear inconsistent with the presence of $Yb^{3+}$ moments inferred from XPS measurements. However, the bulk magnetization should not be interpreted as that of independent free $Yb^{3+}$ local moments. First, the CEF analysis of the ARCS data gives a ground-state Kramers doublet dominated by the $|J_z| = 1/2$ components, with a strongly reduced *c* axis dipolar moment. Second, the magnetization increases with increasing temperature over 250-400 K for both field directions (Extended Data Fig. 13), which is inconsistent with Curie behavior expected for independent or weakly interacting local moments. Third, $Yb^{3+}$ moments in $YbMnBi_2$ are coupled to the Mn magnetic background and to itinerant electrons in a metallic environment; such Yb-Mn coupling, hybridization, and possible valence fluctuations between magnetic $Yb^{3+}$ and nonmagnetic $Yb^{2+}$ can further reduce the apparent bulk magnetic response (52-54). Therefore, the weak magnetization is not evidence against the presence of $Yb^{3+}$ moments. Rather, it reflects the fact that Yb-derived magnetic degrees of freedom in $YbMnBi_2$ are anisotropic, correlated with the Mn sublattice, and embedded in a metallic/intermediate-valence environment. This interpretation is supported by both XPS and the inelastic-neutron-scattering CEF measurements, which establish a significant volume fraction of magnetic $Yb^{3+}$ in the bulk sample (Extended Data Fig. 14).

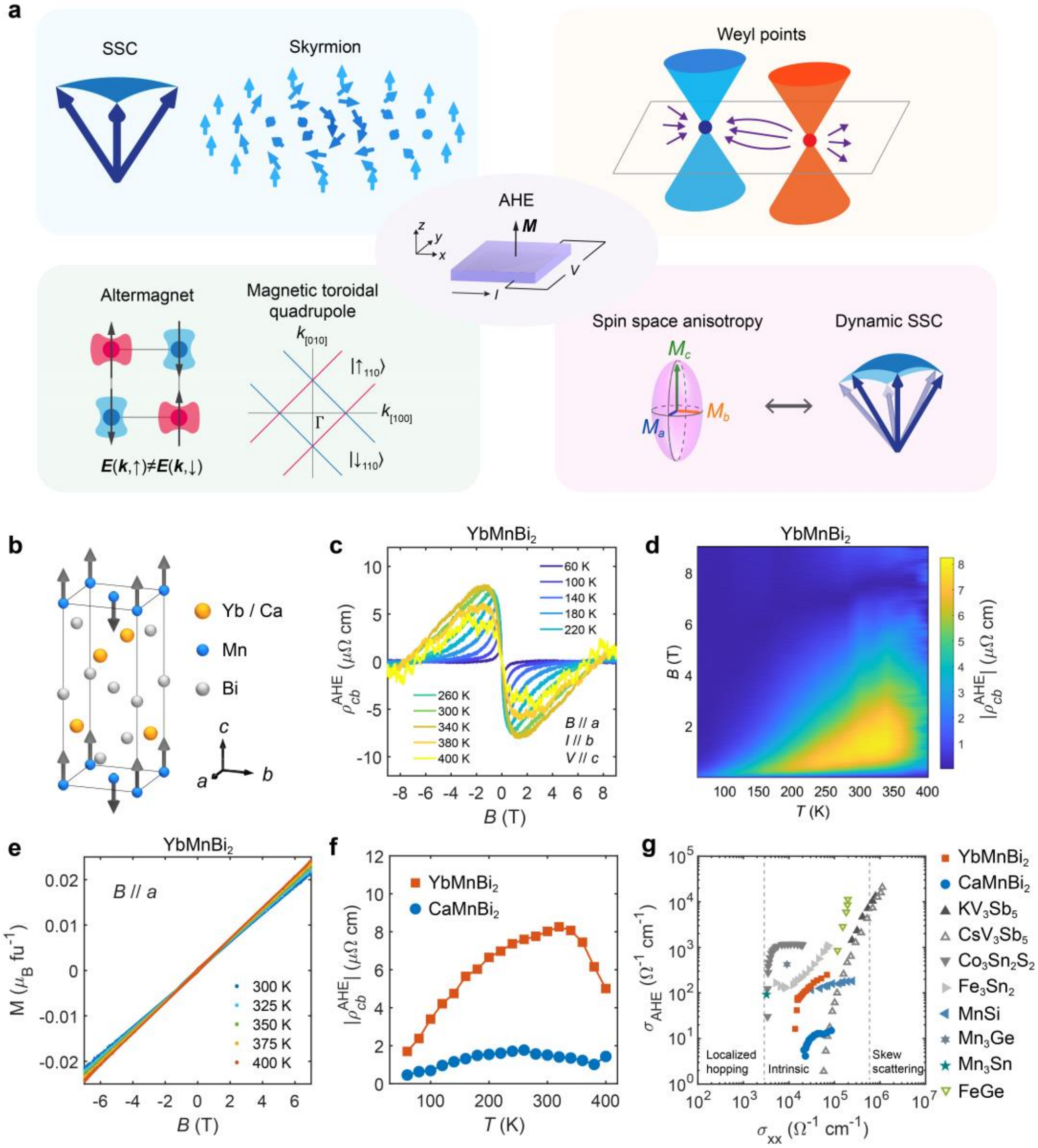

**Figure 1 | Mechanisms of AHE, crystal structure and AHE of $A$MnBi$_2$ . a**, Schematics of mechanisms of AHE, including real-space SSC in non-collinear magnets and skyrmion, Berry curvature in momentum space of Weyl semimetals, altermagnets, and spin nematic state induced dynamic SSC. The pink ellipsoid represents the magnetic responses $M_a$, $M_b$, and $M_c$ measured by neutron LPA along the crystalline *a*-, *b*-, and *c*-axis directions. **b**, Crystal structure of $A$MnBi$_2$ ($A$=Ca,Yb) showing the reported C-type AFM structure(36, 79, 80). **c**, Magnetic field dependence of anomalous Hall resistivity $\rho_{cb}^{\mathrm{AHE}}$ of YbMnBi$_2$. Here, *cb* indicates the measurement of Hall voltage along the *c*-axis and application of current along the *b*-axis. The magnetic field is along the *a*-axis. **d**, The *B*-*T* phase diagram showing the magnetic field and temperature dependence of the absolute value of $\rho_{cb}^{\mathrm{AHE}}$. **e**, Field dependence of the magnetization *M* with magnetic field $\boldsymbol{B}//[100]$ at various temperatures above $T_N$ of YbMnBi$_2$. **f**, Temperature field dependence of anomalous Hall resistivity $\rho_{cb}^{\mathrm{AHE}}$ of $A$MnBi$_2$ ($A$=Ca,Yb). **g**, Anomalous Hall conductivity $\sigma_{AHE}$ versus longitudinal conductivity $\sigma_{xx}$ for $A$MnBi$_2$ and other materials(25, 58, 81-84).

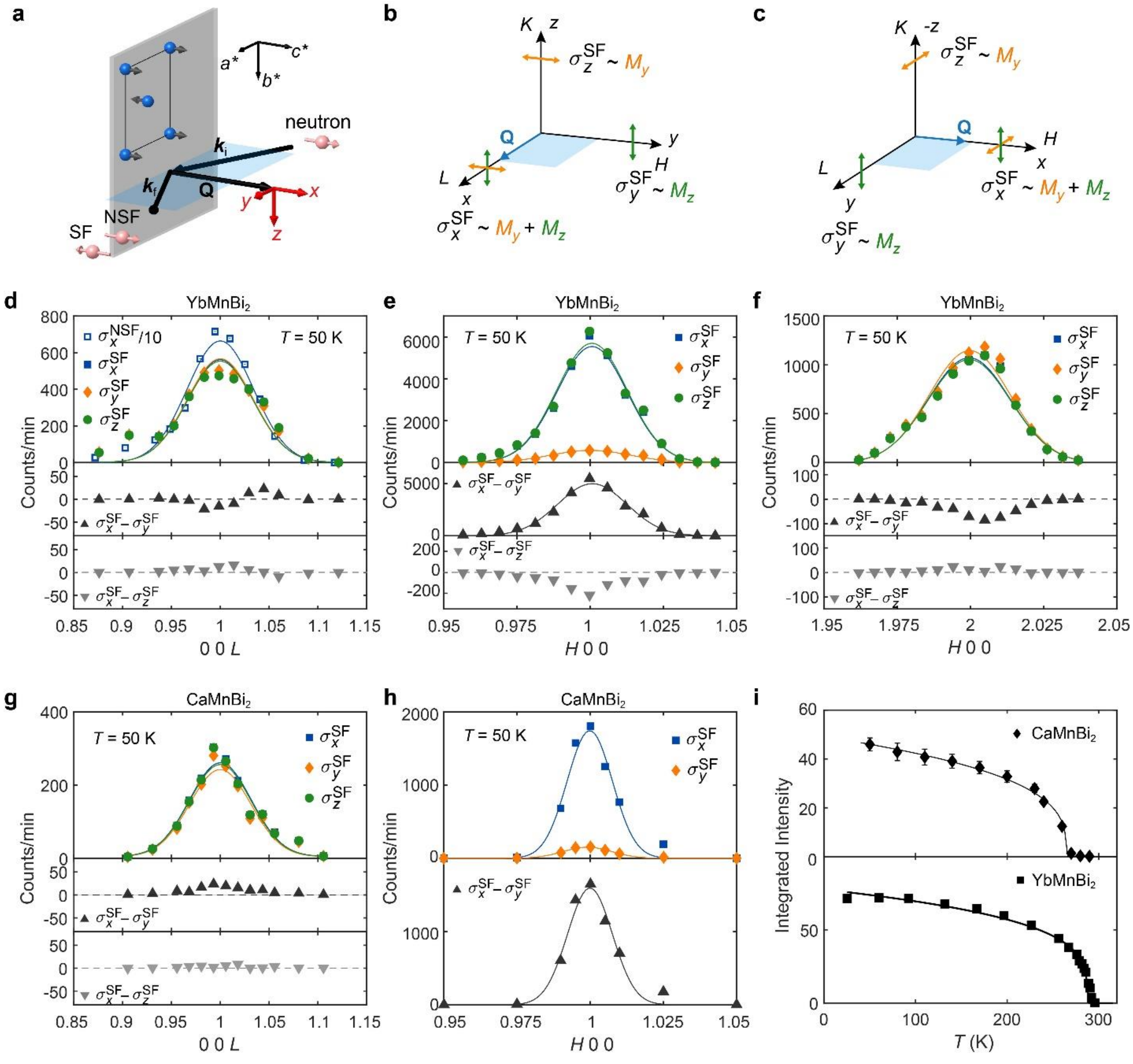

**Figure 2 | Polarized neutron diffraction results of $A$MnBi$_2$. a**, Scattering geometry of polarized neutron scattering experiment in the $[H,0,L]$ scattering plane. Incident neutron spins are polarized along the $\boldsymbol{x}$, $\boldsymbol{y}$, and $\boldsymbol{z}$ directions, corresponding to directions along the $\boldsymbol{Q}$**,** orthogonal to $\boldsymbol{Q}$ within the scattering plane, and perpendicular to the scattering plane, respectively. **b**, SF cross sections $\sigma_{x,y,z}^{\mathrm{SF}}$ and their relationship with $M_{x,y,z}$ for $\boldsymbol{Q}$ along the $[0,0,L]$ direction, where $M_y = M_a$ and $M_z = M_b$. The peak at the (0,0,1) position in $\sigma_{x,y,z}^{\mathrm{SF}}$ is due to imperfect flipping ratio from nuclear Bragg peak. **c**, $\sigma_{x,y,z}^{\mathrm{SF}}$ and their relationship with $M_{x,y,z}$ for $\boldsymbol{Q}$ along the $[H,0,0]$ direction, where $M_y = M_c$ and $M_z = M_b$. **d**, $L$-scans of NSF $\sigma_x^{\mathrm{NSF}}/10$ and SF cross sections $\sigma_{x,y,z}^{\mathrm{SF}}$ across $\boldsymbol{Q} = (0,0,1)$ at 50 K of YbMnBi$_2$. $M_y = \sigma_x^{\mathrm{SF}} - \sigma_y^{\mathrm{SF}} = M_a = 0$ and $M_z = \sigma_x^{\mathrm{SF}} - \sigma_z^{\mathrm{SF}} = M_b = 0$. **e**, $H$-scans of $\sigma_{x,y,z}^{\mathrm{SF}}$ across $\boldsymbol{Q} = (1,0,0)$ at 50 K of YbMnBi$_2$. $M_y = \sigma_x^{\mathrm{SF}} - \sigma_y^{\mathrm{SF}} = M_c$ and $M_z = \sigma_x^{\mathrm{SF}} - \sigma_z^{\mathrm{SF}} = M_b = 0$. **f**, $H$-scans of $\sigma_{x,y,z}^{\mathrm{SF}}$ across $\boldsymbol{Q} = (2,0,0)$ at 50 K of YbMnBi$_2$. $M_y = \sigma_x^{\mathrm{SF}} - \sigma_y^{\mathrm{SF}} =$

$M_c = 0$ and $M_z = \sigma_x^{\mathrm{SF}} - \sigma_z^{\mathrm{SF}} = M_b = 0$. **g**, $L$-scans of SF cross sections $\sigma_{x,y,z}^{\mathrm{SF}}$ across $\boldsymbol{Q} = (0,0,1)$ at 50 K of $CaMnBi_2$. $M_y = \sigma_x^{\mathrm{SF}} - \sigma_y^{\mathrm{SF}} = M_a = 0$ and $M_z = \sigma_x^{\mathrm{SF}} - \sigma_z^{\mathrm{SF}} = M_b = 0$. **h**, $H$-scans of $\sigma_{x,y}^{\mathrm{SF}}$ across $\boldsymbol{Q} = (1,0,0)$ at 50 K of $CaMnBi_2$. $M_y = \sigma_x^{\mathrm{SF}} - \sigma_y^{\mathrm{SF}} = M_c$ and $M_z = \sigma_x^{\mathrm{SF}} - \sigma_z^{\mathrm{SF}} = M_b = 0$. **i**, Magnetic order parameters for AFM measured at $\boldsymbol{Q} = (1,0,0)$ using $\sigma_x^{+-} - \sigma_y^{+-} = M_c$ for $CaMnBi_2$ (top) and $YbMnBi_2$ (bottom). Power-law fits indicate $T_N \approx 270$ K and 290 K for $CaMnBi_2$ and $YbMnBi_2$, respectively.

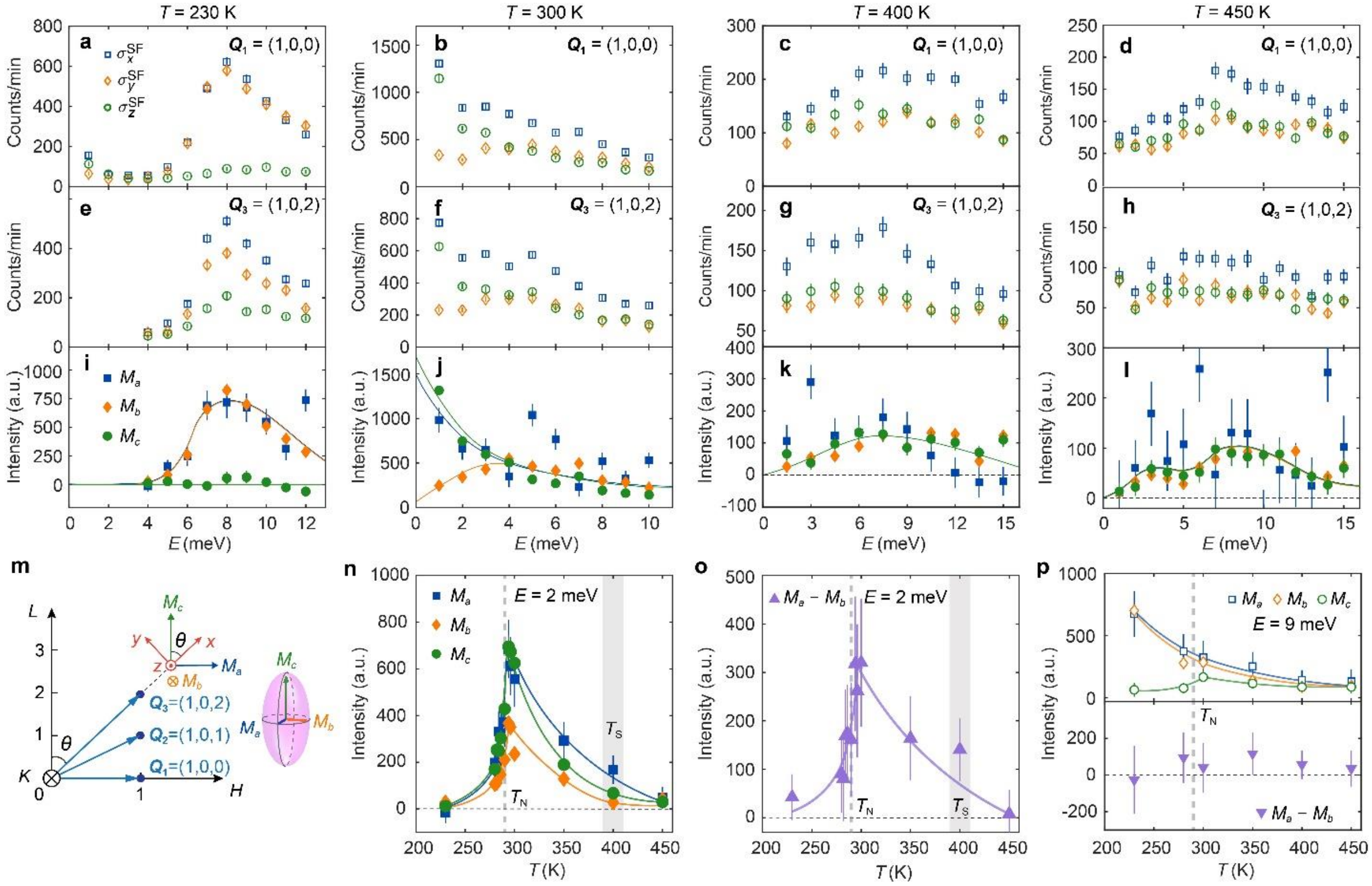

**Figure 3 | Polarized inelastic neutron scattering results of $YbMnBi_2$. l. a-d**, Energy scans of $\sigma_{x,y,z}^{\mathrm{SF}}$ at $\boldsymbol{Q}_1 = (1,0,0)$ measured at $T$ = 230, 300, 400, and 450 K, respectively. **e-h**, Identical scans measured at $\boldsymbol{Q}_3 = (1,0,2)$. **i-l**, Energy dependence of $M_a$, $M_b$, and $M_c$ at $T$ = 230, 300, 400, and 450 K, respectively. **m**, At momentum transfer $\boldsymbol{Q}$, one can probe magnetic response $M_y$ and $M_z$ within the *y*-*z* plane, giving $M_y = M_a \cos^2\theta + M_c \sin^2\theta$ and $M_z = M_b$, where $\theta$ is the angle between $\boldsymbol{Q}$ and $[0,0,L]$. $\boldsymbol{Q}_1 = (1,0,0)$, $\boldsymbol{Q}_2 = (1,0,1)$, $\boldsymbol{Q}_3 = (1,0,2)$ mark the probed wave vectors. By probing two or more equivalent wave vectors with different angle $\theta$, we can conclusively determine $M_a$, $M_b$, and $M_c$. **n**, Temperature dependence of $M_a$, $M_b$, and $M_c$ at $E$ = 2 meV. **o**, Temperature dependence of *ab*-plane magnetic anisotropy ($M_a - M_b$) at $E$ = 2 meV. Vertical shaded area and dashed lines indicate $T_S$ and $T_N$, respectively. **p**, Temperature dependence of $M_a$, $M_b$, and $M_c$, and magnetic anisotropy ($M_a - M_b$) at $E$ = 9 meV.

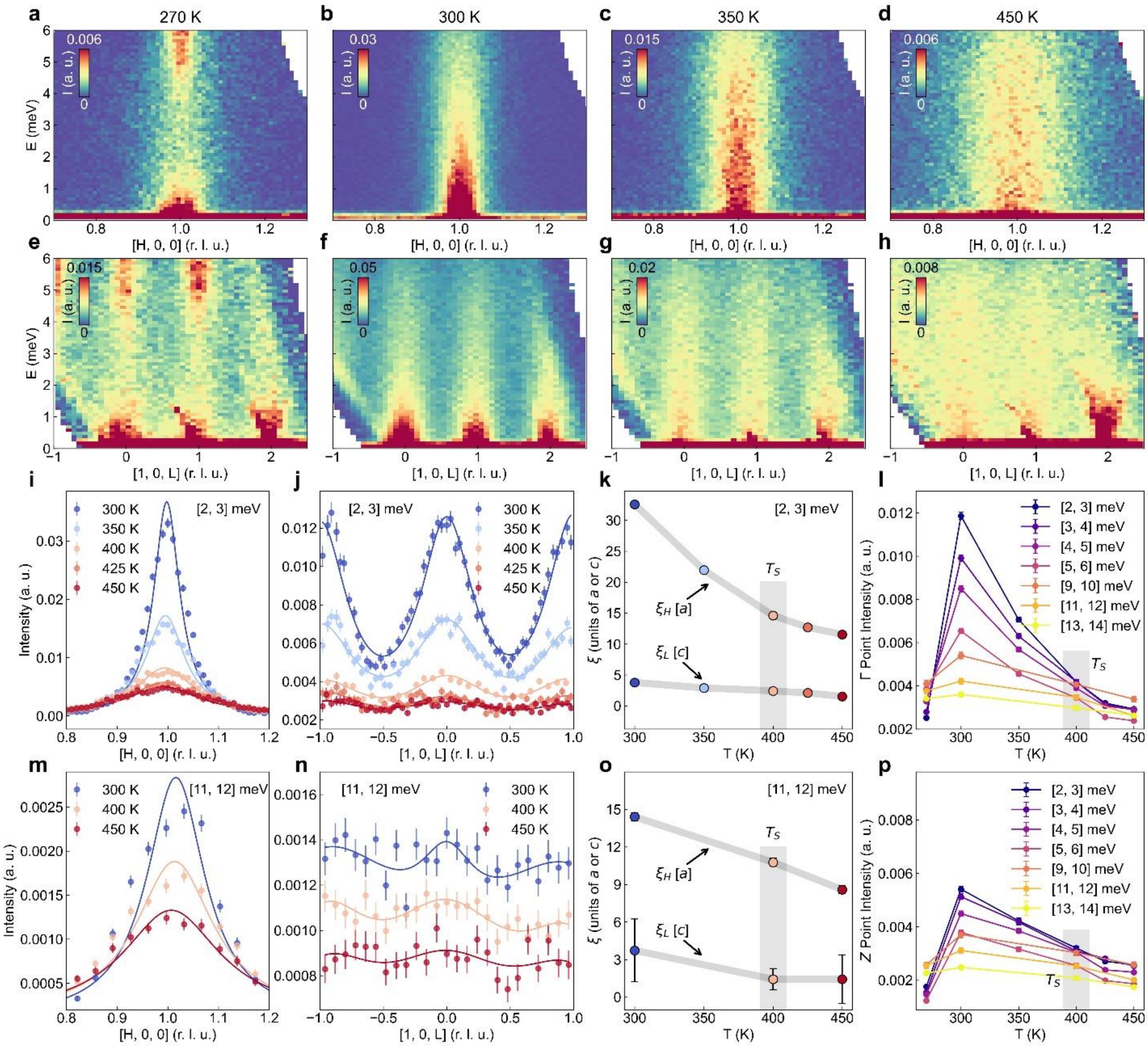

**Figure 4 | Spin dynamics of $YbMnBi_2$ across $T_N$ probed by unpolarized neutrons. a-d**, Inelastic neutron scattering spectra showing spin excitations along the $[H,0,0]$ direction near $\boldsymbol{Q} = (1,0,0)$ at $T = 270$, 300, 350, and 450 K, respectively. **e-h**, Spin excitations along the $[1,0,L]$direction around $\boldsymbol{Q} = (1,0,0)$ measured at the same temperatures. The dispersive low-energy scattering at $(1,0,1)$ and $(1,0,2)$ in **h** is acoustic phonons. **i**, Constant-$E$ cuts integrated over [2, 3] meV along the $[H,0,0]$ direction at different temperatures. The solid lines indicate the Lorentzian fit to extract the in-plane correlation length. **j,** Similar cuts of spin excitation along the $[1,0,L]$ direction, fitted with three-peak Lorentzian to extract out-of-plane correlation length. **m-n,** Similar cuts at $E = 11.5 \pm 0.5$ meV. **k**,**o**, Temperature dependence of the correlation lengths expressed in lattice units, with $\xi_H$ referenced to the in-plane lattice constant $a$, and $\xi_L$ to the out-of-plane lattice constant $c$. obtained from the fitting results in **i**,**j**,**m**,**n**. The shaded line marks the nematic transition temperature $T_S$, below which spin-space anisotropy emerges. **l,** Temperature dependence of the scattering intensity $S(\boldsymbol{Q},E)$ at the Γ point, integrated over multiple energy

windows. **p,** Temperature dependence of the spin excitation intensity at the $Z$ point, extracted from constant-$\boldsymbol{Q}$ cuts near $\boldsymbol{Q} = (1,0,0.5)$.

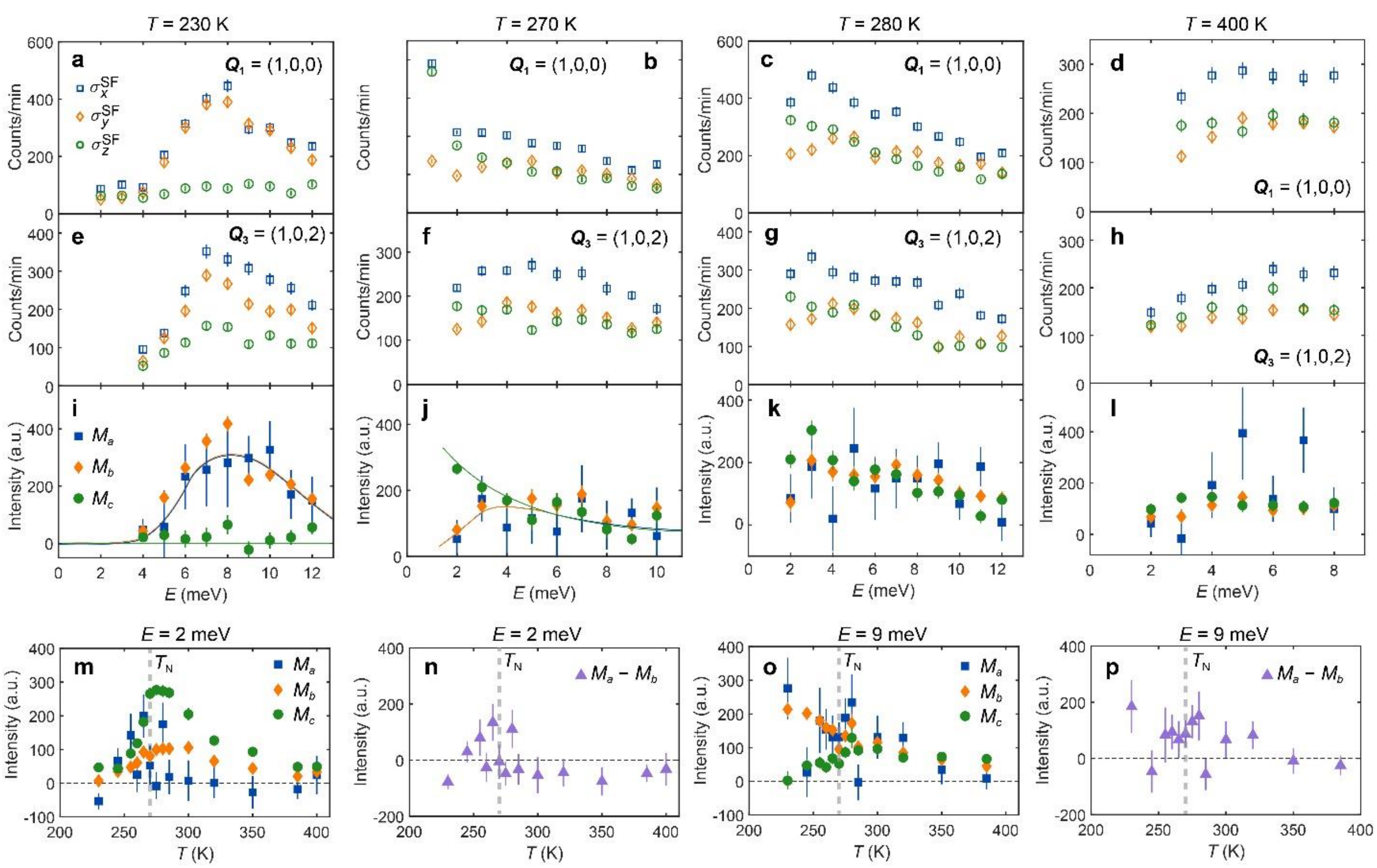

**Figure 5 | Polarized inelastic neutron scattering results of $CaMnBi_2$. a-d**, Energy scans of $\sigma_{x,y,z}^{\mathrm{SF}}$ at $\boldsymbol{Q}_1 = (1,0,0)$ measured at $T = 230$, 270, 280, and 400 K, respectively. **e-h**, Identical scans measured at $\boldsymbol{Q}_3 = (1,0,2)$. **i-l**, Energy dependence of $M_a$, $M_b$, and $M_c$ at $T = 230$, 270, 280, and 400 K, respectively. **m**, Temperature dependence of $M_a$, $M_b$, and $M_c$ at $E = 2$ meV. **n**, Temperature dependence of $ab$-plane magnetic anisotropy ($M_a - M_b$) at $E = 2$ meV. **o-p**, Similar data at $E = 9$ meV. Dashed lines indicate $T_N$.

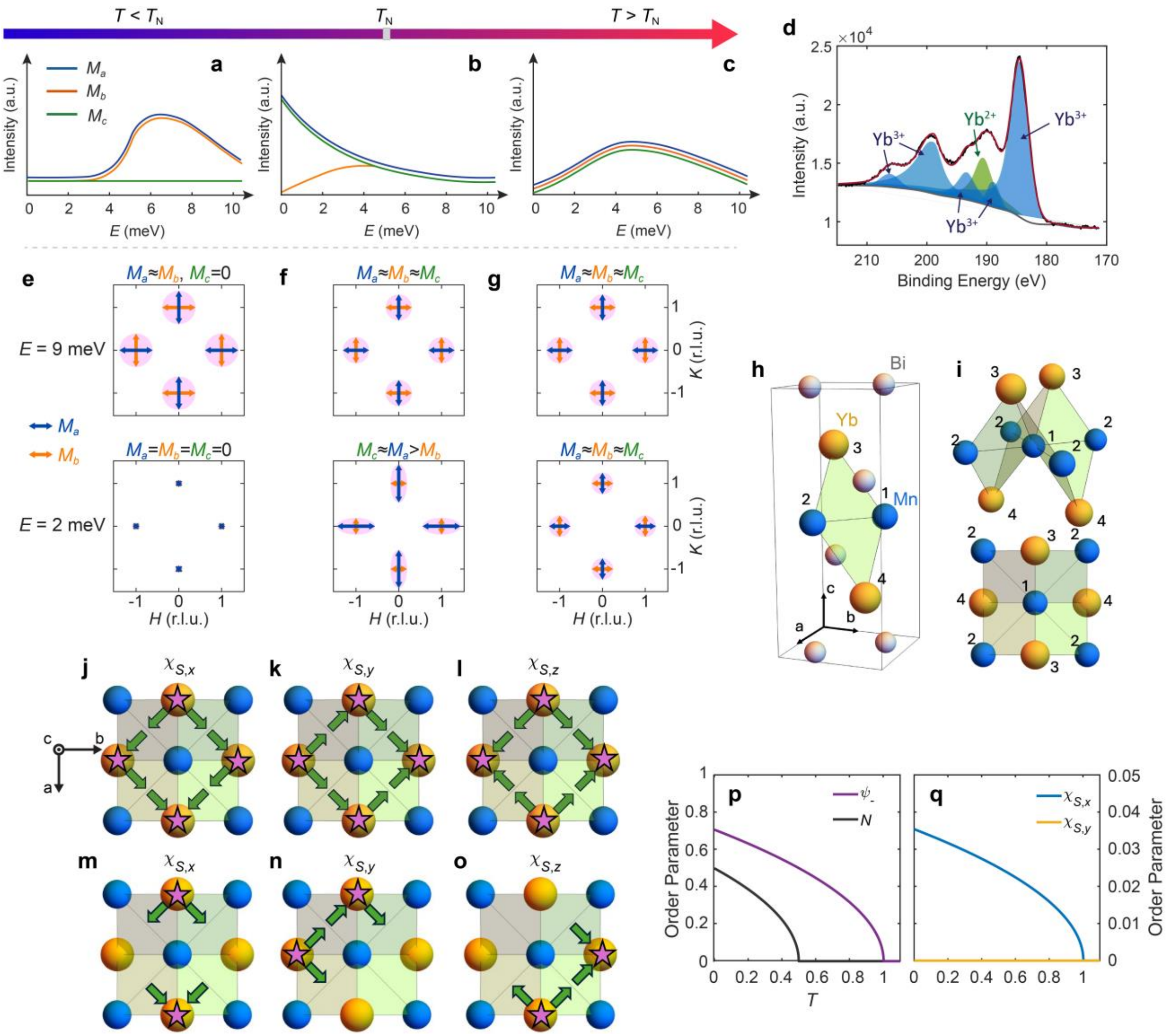

**Figure 6, Experimental and theoretical understanding of magnetism and SSC in YbMnBi₂.** **a**,**b**,**c**, Schematics of energy dependence of $M_a$, $M_b$, and $M_c$ at temperatures below $T_N$, around $T_N$, and well above $T_N$, respectively. **d**, XPS result of $YbMnBi_2$ indicates that majority of the Yb ions are in the trivalent states. **e**,**f**,**g**, Schematics of the magnetic responses $M_a$, $M_b$, and $M_c$ at $E$ = 9 meV (top row) and $E$ = 2 meV (bottom row). At temperatures well below $T_N$, $M_a$ and $M_b$ are isotropic with $M_c$ = 0 at $E$ = 9 meV, and all three responses are fully gapped out at $E$ = 2 meV. Around $T_N$,the magnetic responses at $E$ = 9 meV remain isotropic, whereas at $E$ = 2 meV it develops in-plane anisotropy ($M_a > M_b$). Way above $T_N$, all three responses become isotropic ($M_a \approx M_b \approx M_c$) at both energies. **h**, Unit cell of $A$MnBi₂, showing Bi (grey), Mn (blue, 1-2), and $A$ = Yb, Ca (gold, 3-4), If $A$ is the magnetic $Yb^{3+}$, the green-shaded Mn-Yb-Mn triangle is formed and the finite SSC emerges. **i**, The eight triangles surrounding each Mn ion from (top panel) side view and (bottom panel) top view. The configuration of SSC of each triangle determines the direction of AHE. **j**-**l**, The configuration of SSC $\boldsymbol{\chi}_S = (\chi_{S,x}, \chi_{S,y}, \chi_{S,z})$ with full $Yb^{3+}$ occupancy from the top view. **m-o**, The configuration of SSC with 50% $Yb^{3+}$ occupancy, where stars mark

the $Yb^{3+}$ sites. **p**,**q**, Temperature dependence of order parameters at finite $\boldsymbol{Q}$, using $T_S = 1$, $T_A = 0.5$, $\alpha(Q \neq 0) = \alpha_4 = \beta_4 = \eta_1 = \eta_2 = 1$, $g_2 = -0.1$, and $\boldsymbol{B} = (1,0,0)$.

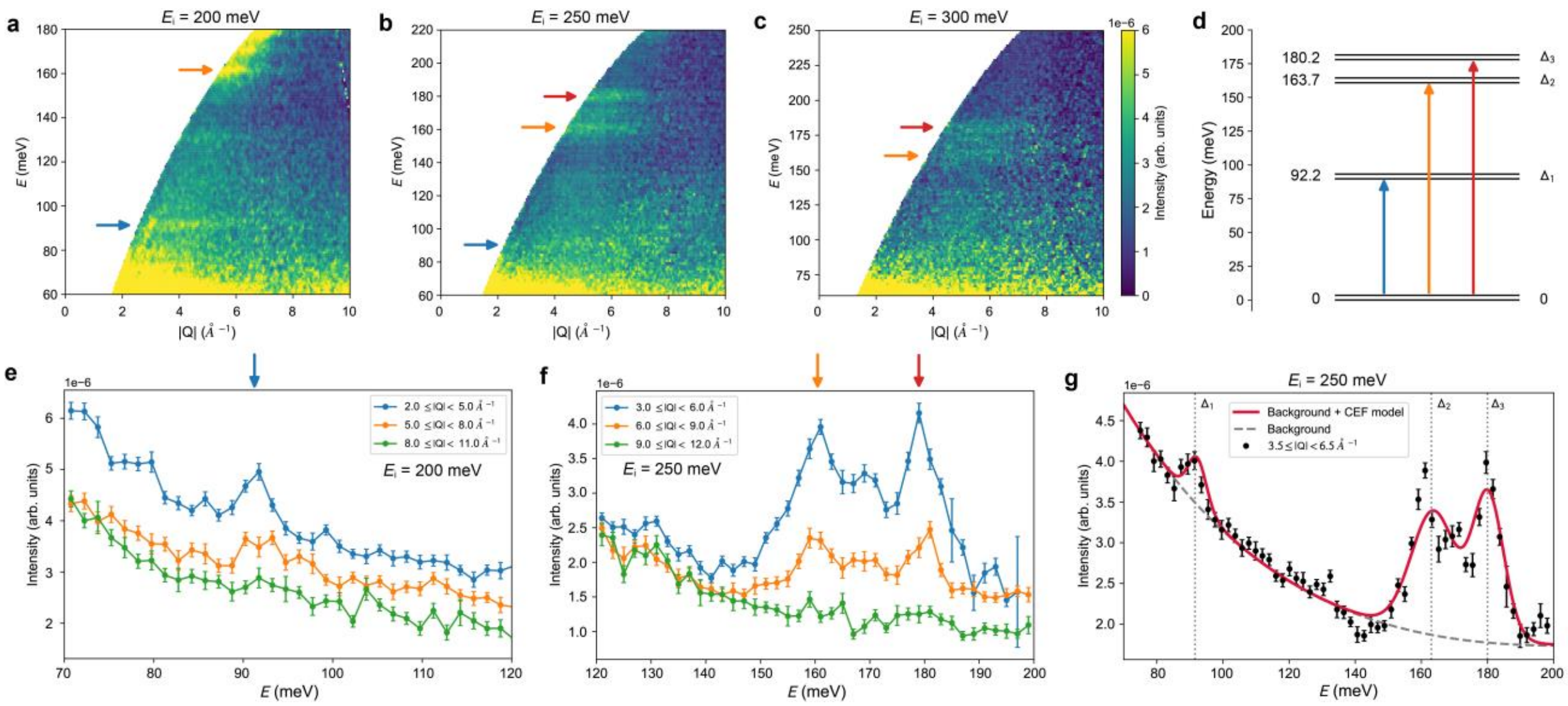

**Figure 7 Inelastic neutron scattering evidence for $Yb^{3+}$ crystal-electric-field excitations in $YbMnBi_2$.** **a**-**c**, Inelastic neutron scattering intensity maps measured with incident energies $E_i = 200, 250, 300$ meV and a frequency of 600 Hz. All three panels share the same intensity scale shown by the color bar in **c**. Arrows indicate candidate crystal electric field excitations. **d**, CEF level scheme for $Yb^{3+}$ in $YbMnBi_2$, showing transitions at $\Delta_1 = 92.2$, $\Delta_2 = 163.7$, and $\Delta_3 = 180.2$ meV. **e**,**f**, Constant-$|\boldsymbol{Q}|$ cuts showing nondispersive excitations whose intensities decrease with increasing $|\boldsymbol{Q}|$, consistent with magnetic CEF scattering. **g**, Constant-$|\boldsymbol{Q}|$ cut for $3.5 \leq |Q| < 6.5$ Å$^{-1}$ compared with the calculated CEF spectrum. The CEF parameters were determined using the average peak energies and normalized integrated spectral weights extracted from the $E_i = 250$ and 300 meV datasets, as shown in Extended Data Fig. 14. The solid red curve represents calculated CEF spectrum plus an empirical background, while the dashed gray curve shows the background contribution. The vertical dotted lines mark the CEF transition energies used in the calculation.

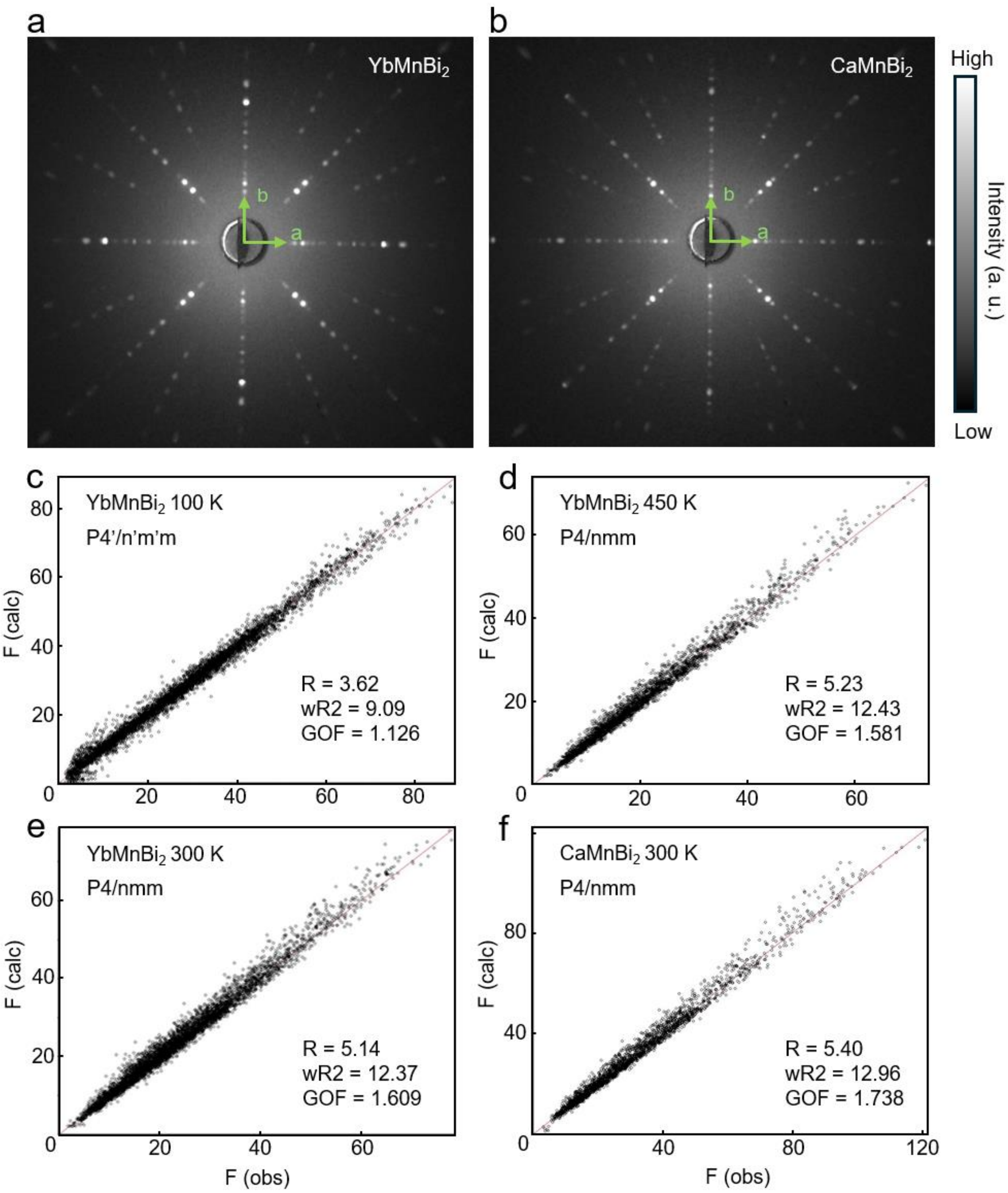

**Extended Data Fig. 1 | X-ray Laue image, Single crystal neutron diffraction structure refinement of (Yb, Ca) $MnBi_2$.** **a**,**b**, X-ray backscattering laue image of $YbMnBi_2$ and $CaMnBi_2$ showing 001 plane. **c-e**, Single-crystal neutron diffraction structure refinements of $YbMnBi_2$ at (**c**) 100 K, (**d**) 450 K, and (**e**) 300 K, using the corresponding space group. **f**, Single-crystal neutron diffraction structure refinements of $CaMnBi_2$ at 300 K, performed using the corresponding space group.

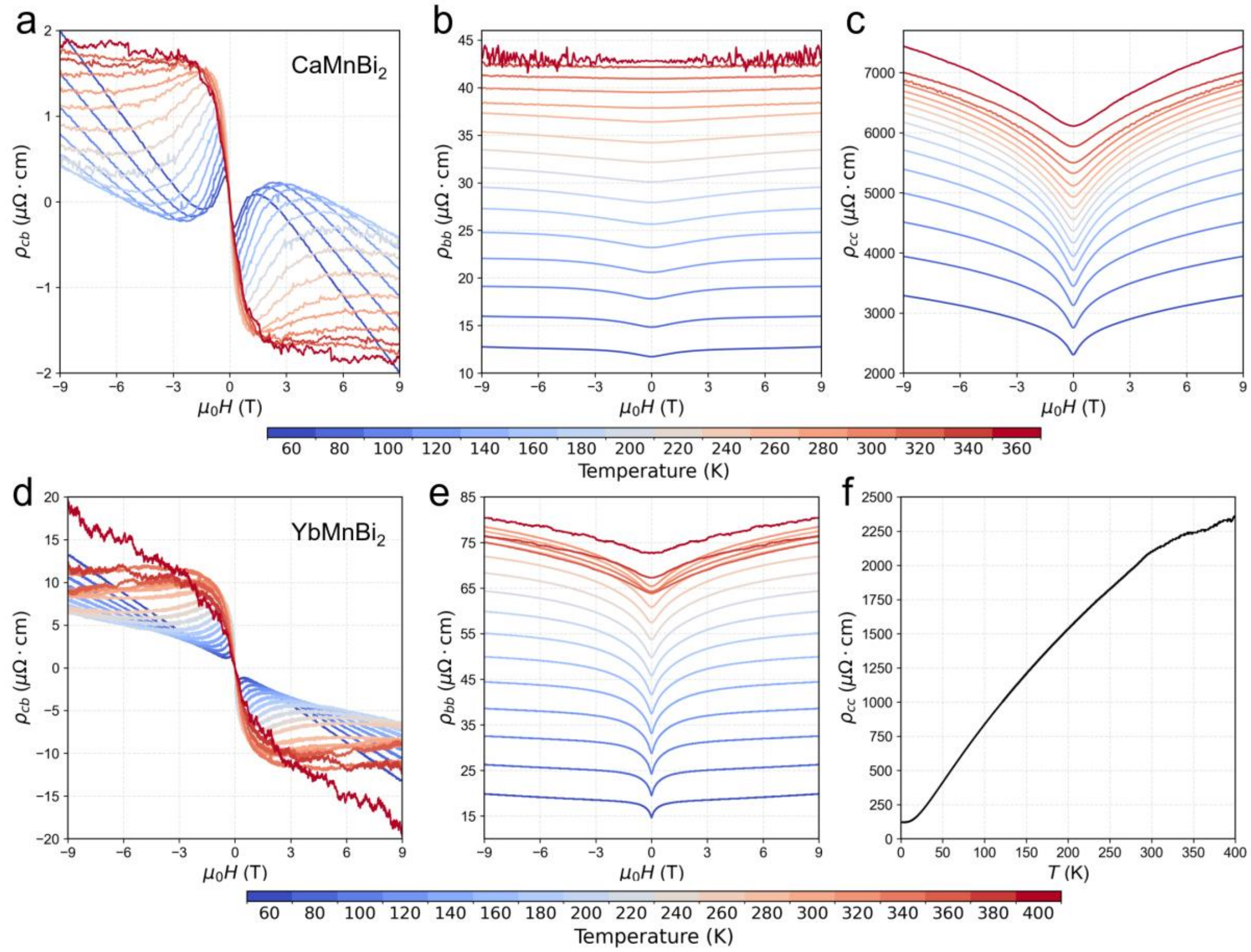

**Extended Data Fig. 2 | Magnetic field dependence of Hall resistivity and Magneto resistivity of (a,b,c) $CaMnBi_2$ and (d,e,f) $YbMnBi_2$.** $\rho_{cb}$ measurements were performed with the magnetic field $B$ applied along $a\|[1,0,0]$, current along $b\|[0,1,0]$, and Hall voltage along $c\|[0,0,1]$. $\rho_{bb}$ and $\rho_{cc}$ measurements were performed with the magnetic field B applied along $a\|[1,0,0]$, current and voltage along $b\|[0,1,0]$ and $c\|[0,0,1]$, respectively.

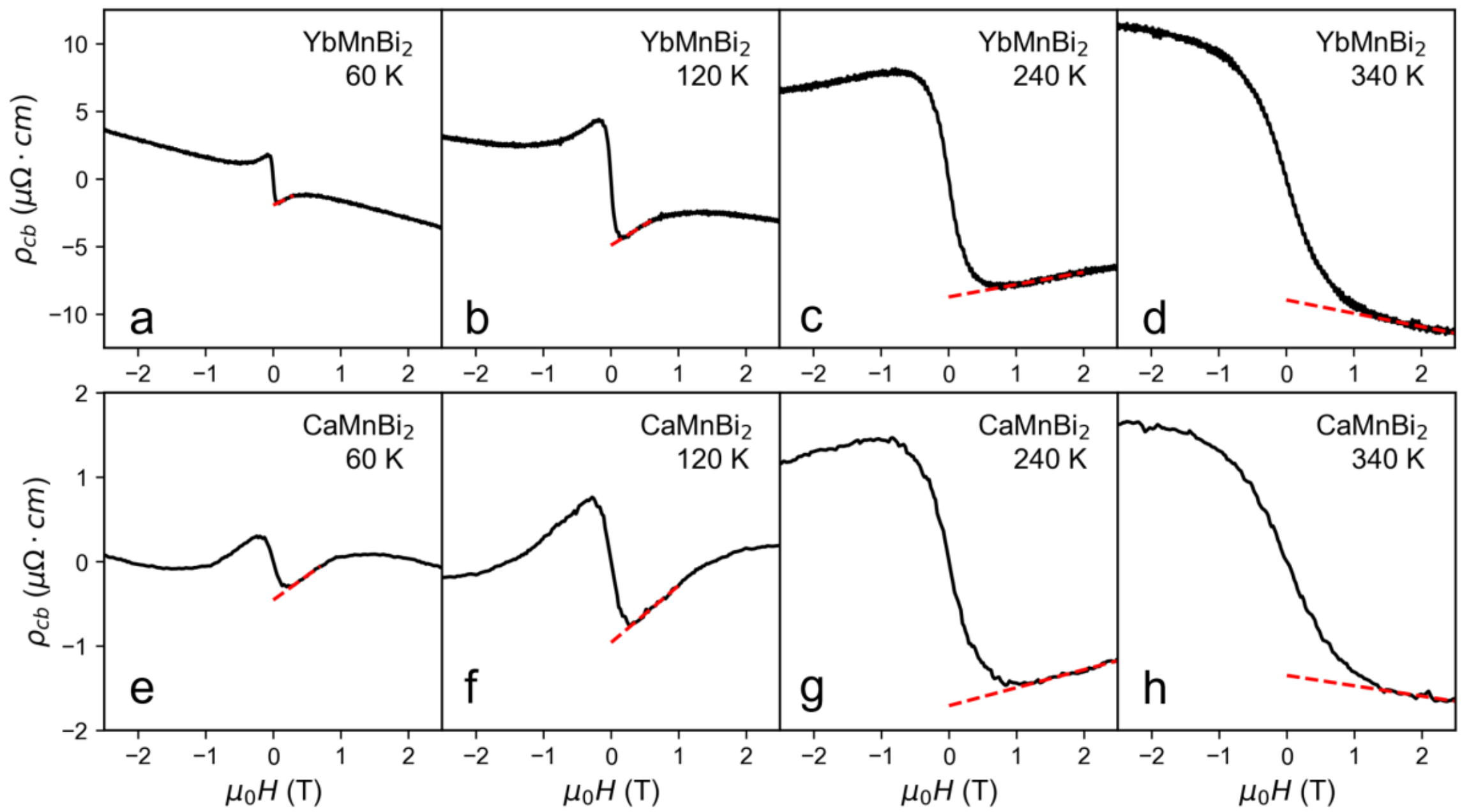

**Extended Data Fig. 3 | Hall resistivity of (a-d) $YbMnBi_2$ and (e-h) $CaMnBi_2$ at different temperatures.** The zoomed-in plots highlight the Hall resistivity data of (Yb, Ca) $MnBi_2$ from -2.5 to 2.5 T at different temperatures. The red dashed line indicates the linear extrapolation of the data within a defined linear range. In the Hall response of (Yb, Ca) $MnBi_2$, a sharp low-field upturn is observed, which cannot be attributed to the ordinary Hall effect or a multi-band effect, as noted in previous studies of this compound family(85). Instead, this feature is identified as an anomalous Hall effect (AHE). The red curve represents a linear fit in the low-field region, with the absolute value of the intercept providing an estimation of the AHE signal.

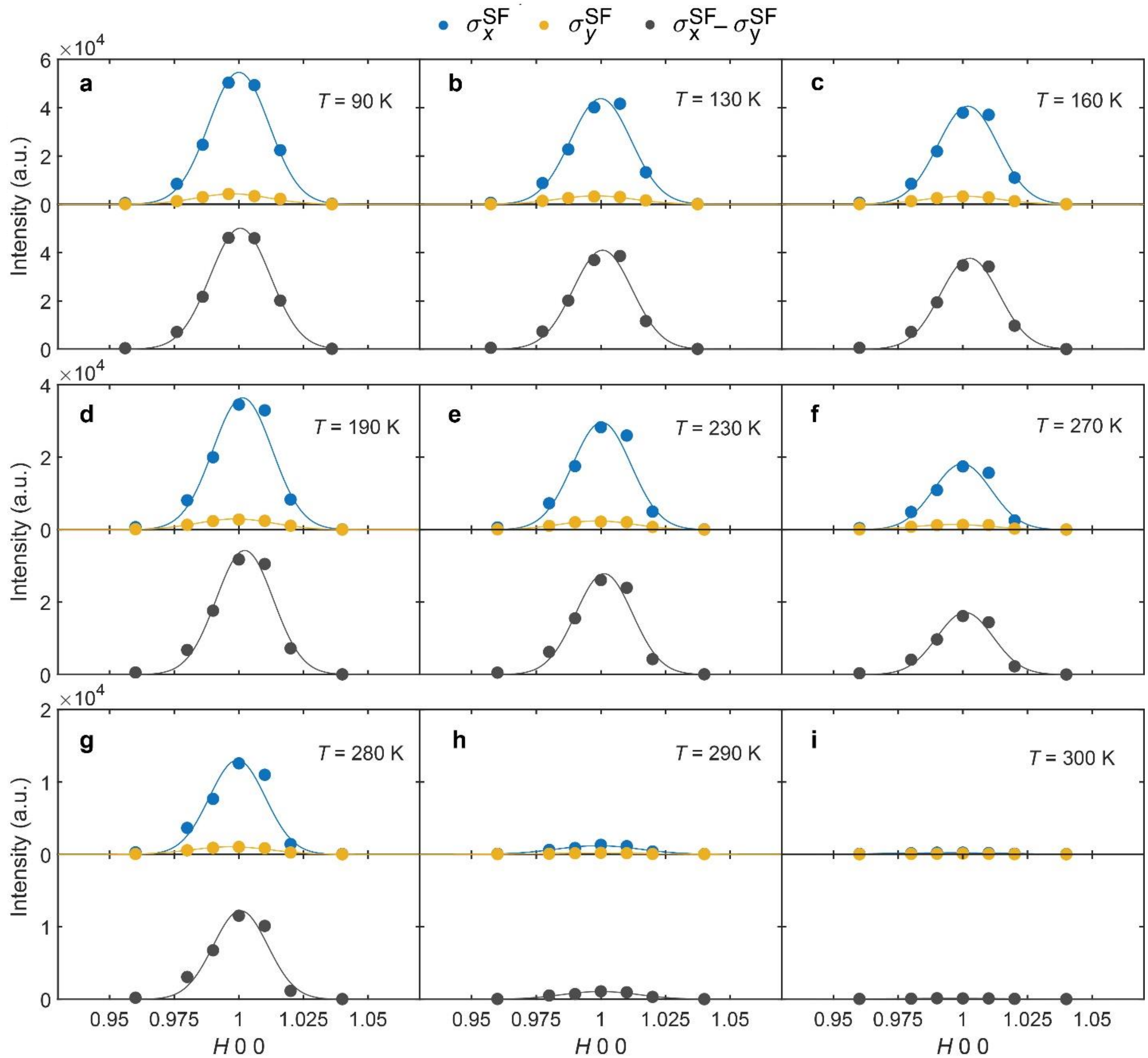

**Extended Data Fig. 4 | *H*-scans of $\sigma_x^{\text{SF}}$ and $\sigma_y^{\text{SF}}$ of $\text{YbMnBi}_2$ at various temperatures.** **a-i**, *H*-scans of $\sigma_x^{\text{SF}}$ and $\sigma_y^{\text{SF}}$ of $\text{YbMnBi}_2$ at various temperatures. The temperature dependence of integrated intensity of $\sigma_x^{\text{SF}} - \sigma_y^{\text{SF}}$ corresponds to the AFM order parameter in Fig. 2i.

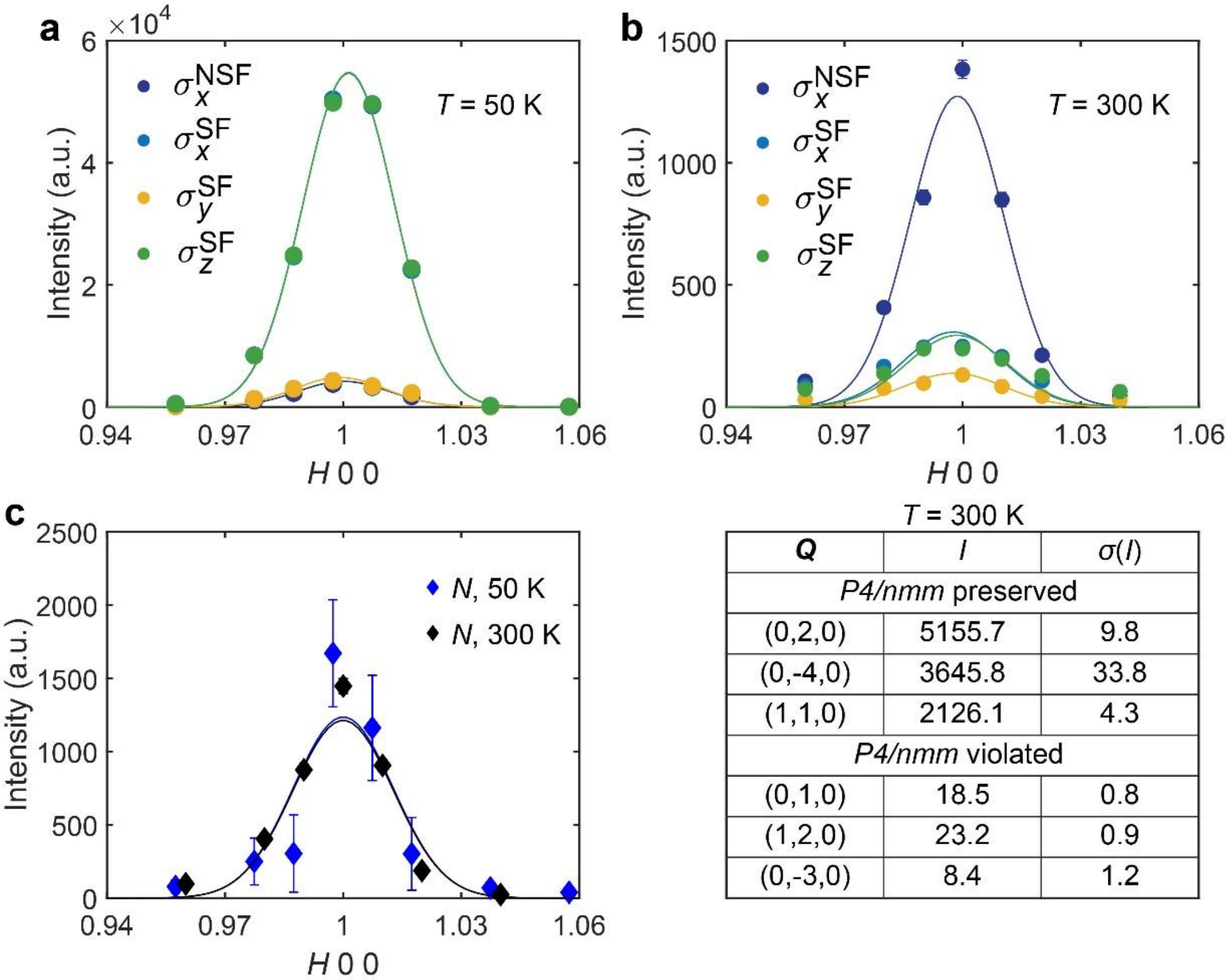

| *Q* | *I* | σ(*I*) |
|---|---|---|
| *P4/nmm* preserved | | |
| (0,2,0) | 5155.7 | 9.8 |
| (0,-4,0) | 3645.8 | 33.8 |
| (1,1,0) | 2126.1 | 4.3 |
| *P4/nmm* violated | | |
| (0,1,0) | 18.5 | 0.8 |
| (1,2,0) | 23.2 | 0.9 |
| (0,-3,0) | 8.4 | 1.2 |

**Extended Data Fig. 5 | *H*-scans of $\sigma^{SF}_{x,y,z}$ and $\sigma^{NSF}_{x}$ of YbMnBi$_2$ at 50 K and 300 K. a**,**b**, *H*-scans of $\sigma^{SF}_{x,y,z}$ and $\sigma^{NSF}_{x}$ of YbMnBi$_2$ at 50 K and 300 K, respectively. **c**, *H*-scans of nuclear scattering *N* deduced from cross sections in **a** and **b**. $\sigma^{SF}_{x} \sim \sigma^{SF}_{z} > \sigma^{SF}_{y}$ indicate magnetic scattering from the AFM *c*-axis moment. Despite being forbidden by the reported space group *P4/nmm*, non-spin-flip intensity is observed for $\sigma^{++}_{x}$ at (1,0,0). The similar intensity of nuclear scattering at 50 K and 300 K suggests the nuclear intensity at $\boldsymbol{Q} = (1,0,0)$ may come from small defects and is temperature independent. A comparison of the integrated scattering for a few fundamental and forbidden peaks from TOPAZ data is shown in the Table.

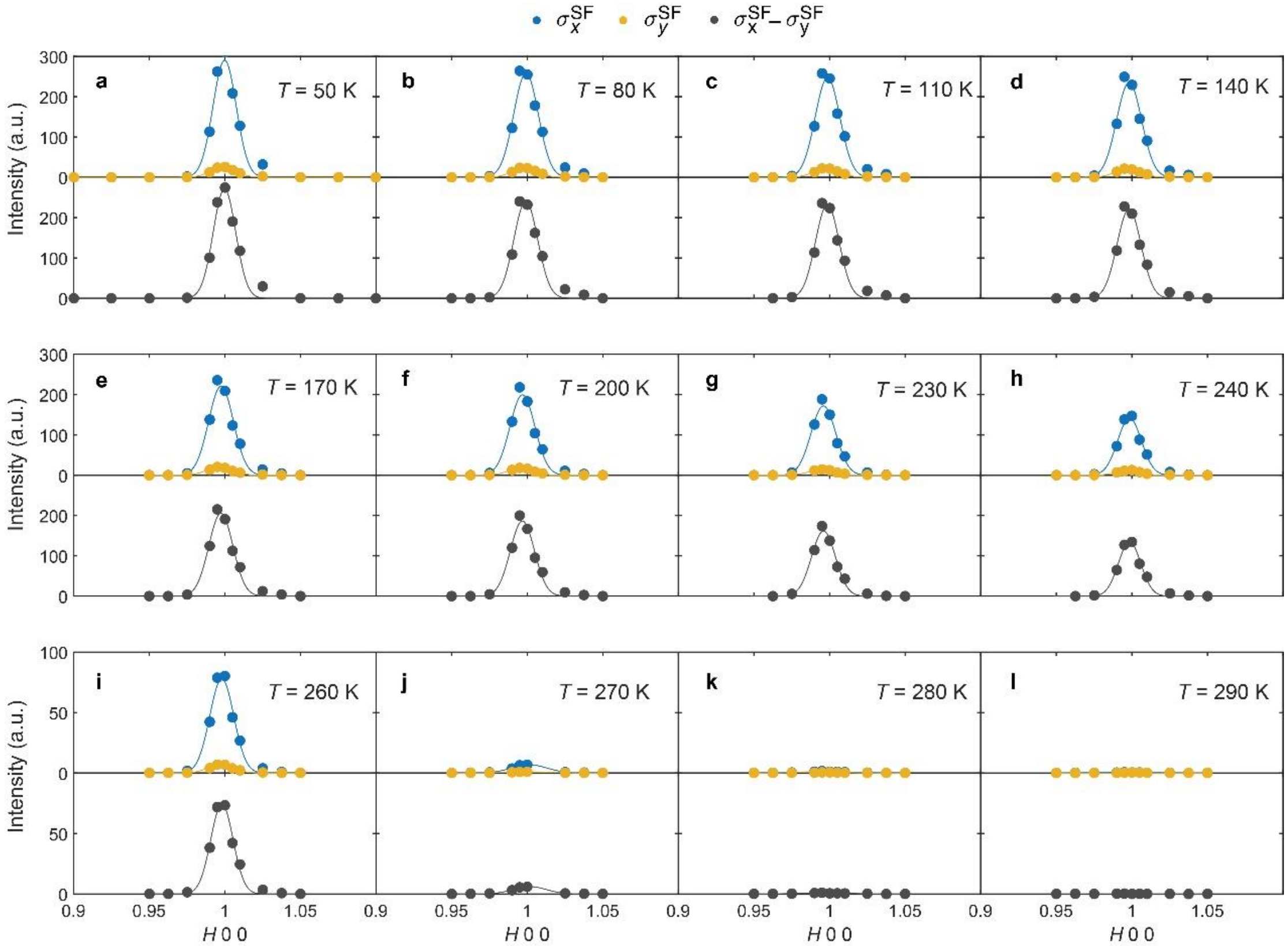

**Extended Data Fig. 6 | *H*-scans of $\sigma_x^{SF}$ and $\sigma_y^{SF}$ of $CaMnBi_2$ at various temperatures. a-l**, *H*-scans of $\sigma_x^{SF}$ and $\sigma_y^{SF}$ of $CaMnBi_2$ at various temperatures. The temperature dependence of integrated intensity of $\sigma_x^{SF} - \sigma_y^{SF}$ corresponds to the AFM order parameter in Fig. 2i.

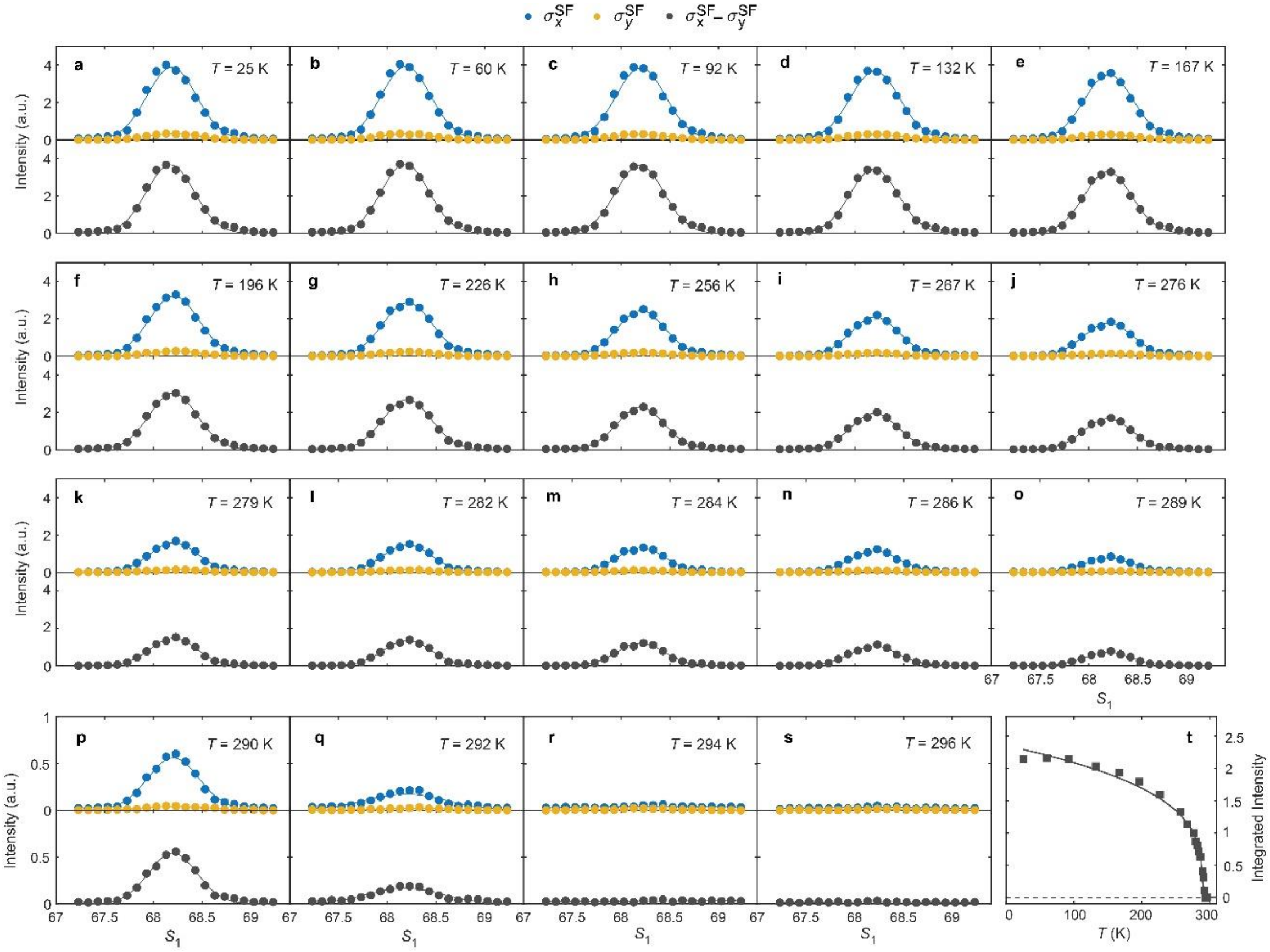

**Extended Data Fig. 7 | Rocking scans of $\sigma_x^{SF}$ and $\sigma_y^{SF}$ of $YbMnBi_2$ sample #2 at $\boldsymbol{Q} = (1,0,0)$ at various temperatures.** a-s, Rocking scans of $\sigma_x^{SF}$ and $\sigma_y^{SF}$ of $YbMnBi_2$ at $\boldsymbol{Q} = (1,0,0)$ at various temperatures. **t**, Temperature dependence of integrated intensity of $\sigma_x^{SF} - \sigma_y^{SF}$. Power-law fit indicates $T_N \approx 290$ K.

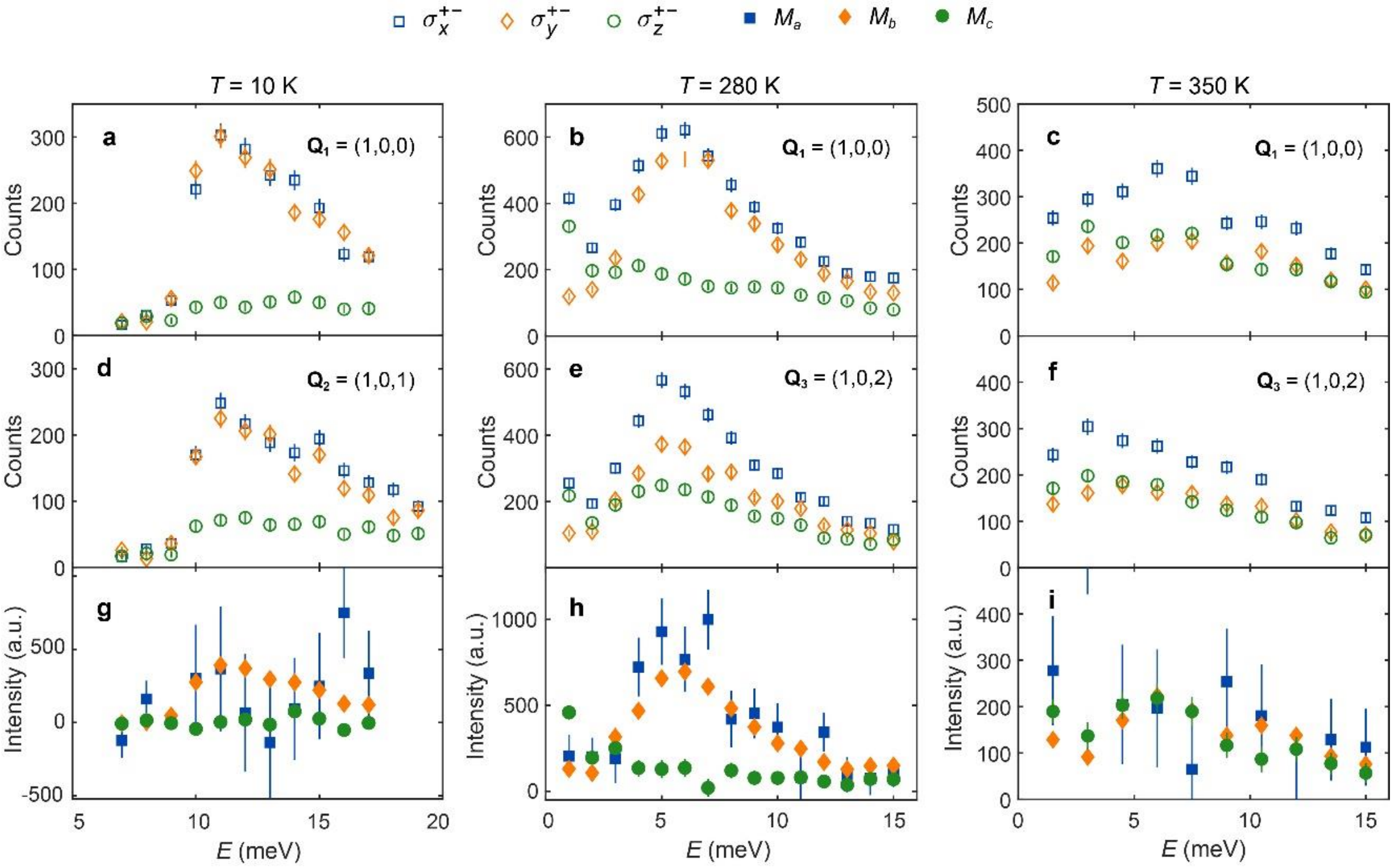

**Extended Data Fig. 8 | Polarized inelastic neutron scattering results of YbMnBi₂. a-c**, Energy scans of $\sigma_{x,y,z}^{\mathrm{SF}}$ at $\boldsymbol{Q}_1 = (1,0,0)$ measured at $T$ = 10, 280, and 350 K, respectively. **d**, Energy scans of $\sigma_{x,y,z}^{\mathrm{SF}}$ at $\boldsymbol{Q}_2 = (1,0,1)$ at $T$ = 10 K. **e**,**f**, Energy scans of $\sigma_{x,y,z}^{\mathrm{SF}}$ at $\boldsymbol{Q}_3 = (1,0,2)$ at $T$ = 280 and 350K, respectively. **g-i**, Energy dependence of $M_a$, $M_b$, and $M_c$ at $T$ = 10, 280, and 350 K, respectively.

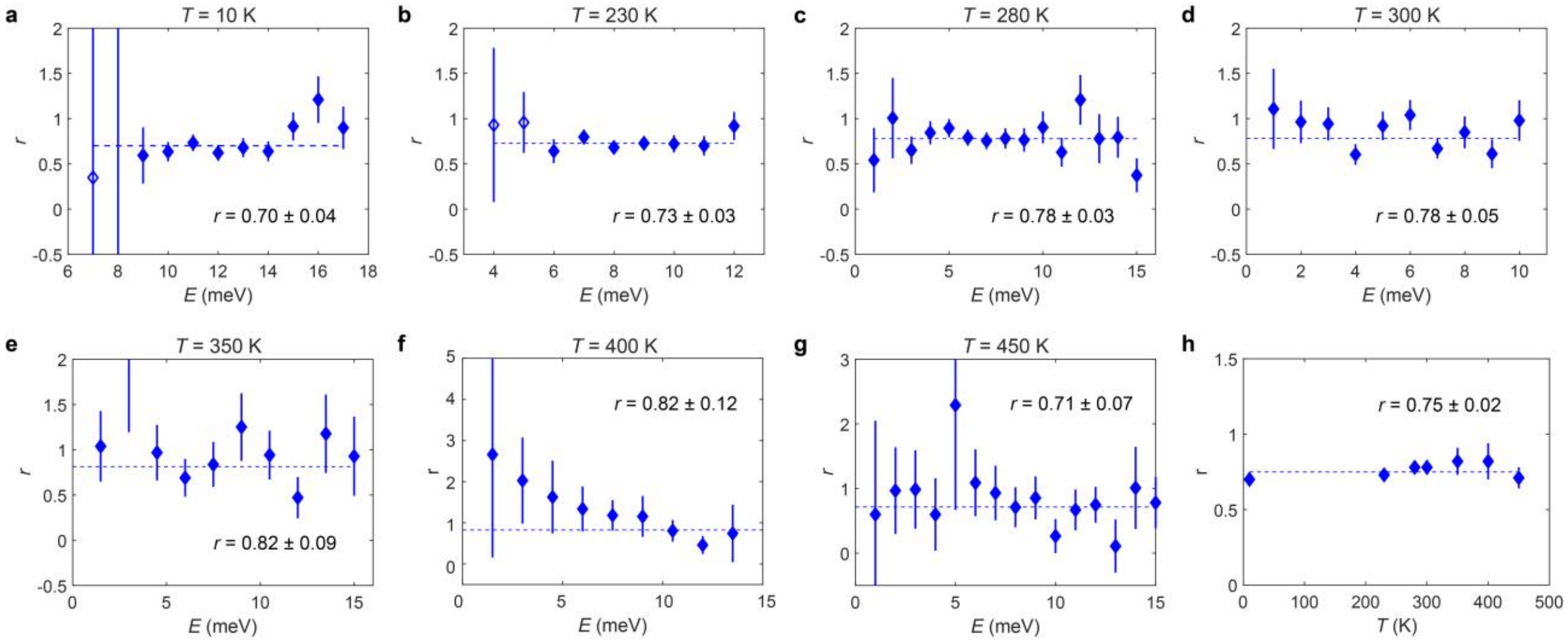

**Extended Data Fig. 9 | Energy and temperature dependence of intensity ratio factor *r* for YbMnBi₂. a-g**, Intensity ratio factor r obtained for energy scans of $\sigma_{x,y,z}^{\mathrm{SF}}$ at $\boldsymbol{Q}_1 = (1,0,0)$ and $\boldsymbol{Q}_3 = (1,0,2)$ measured at $T$ = 10, 230, 280, 300, 350, 400, and 450 K, respectively. The dashed lines are constant (flat line) fits, and the values of *r* shown in each panel are the results of these

fits. Open symbols are excluded from the fits. **h**, Temperature dependence of *r*, where the dash line is a flat line fit.

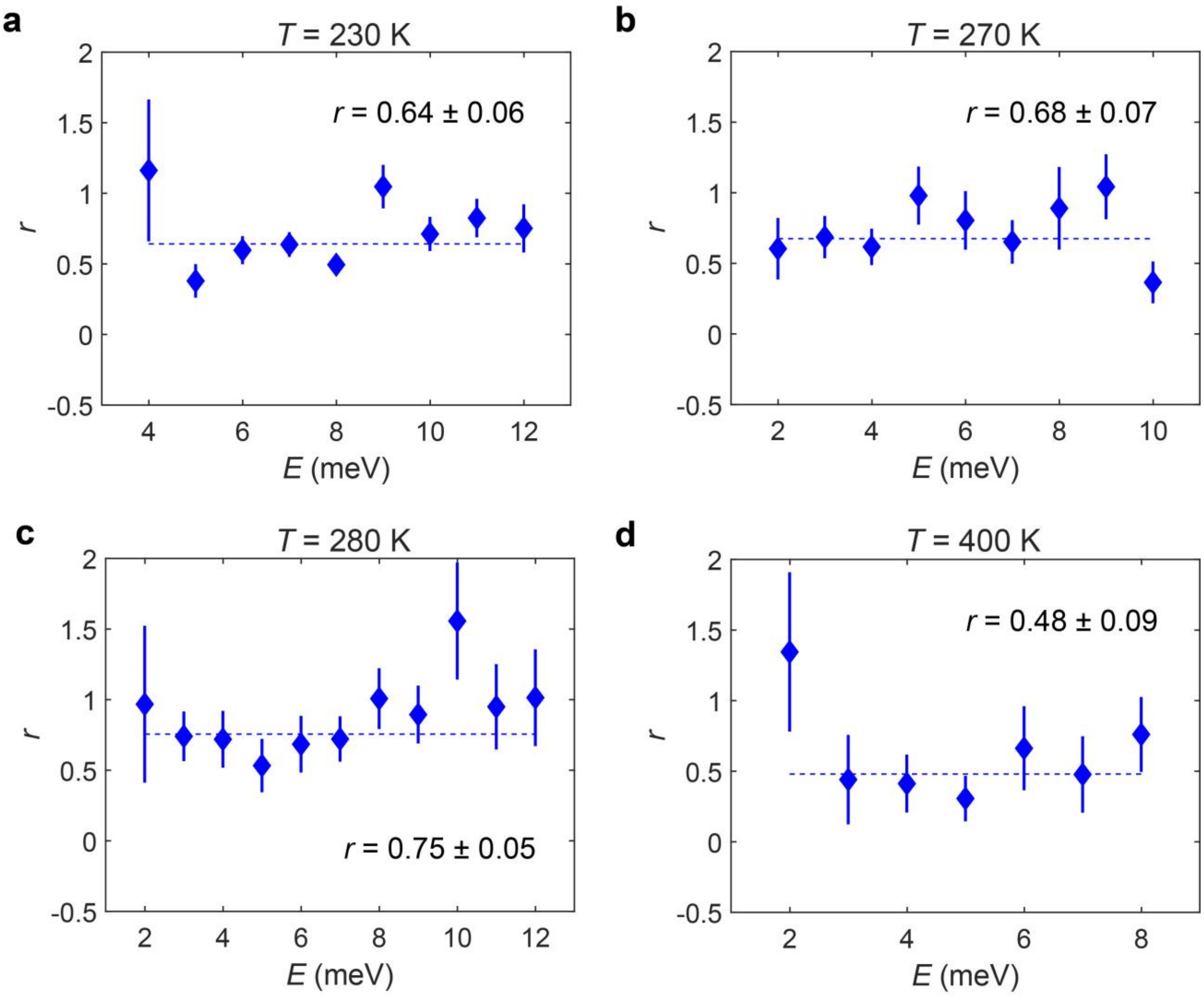

**Extended Data Fig. 10 | Energy and temperature dependence of intensity ratio factor *r* for CaMnBi₂. a-g**, Intensity ratio factor r obtained for energy scans of $\sigma_{x,y,z}^{\mathrm{SF}}$ at $\boldsymbol{Q}_1 = (1,0,0)$ and $\boldsymbol{Q}_3 = (1,0,2)$ measured at $T$ = 230, 270, 280, and 400 K, respectively. The dashed lines are constant (flat line) fits, and the values of *r* shown in each panel are the results of these fits.

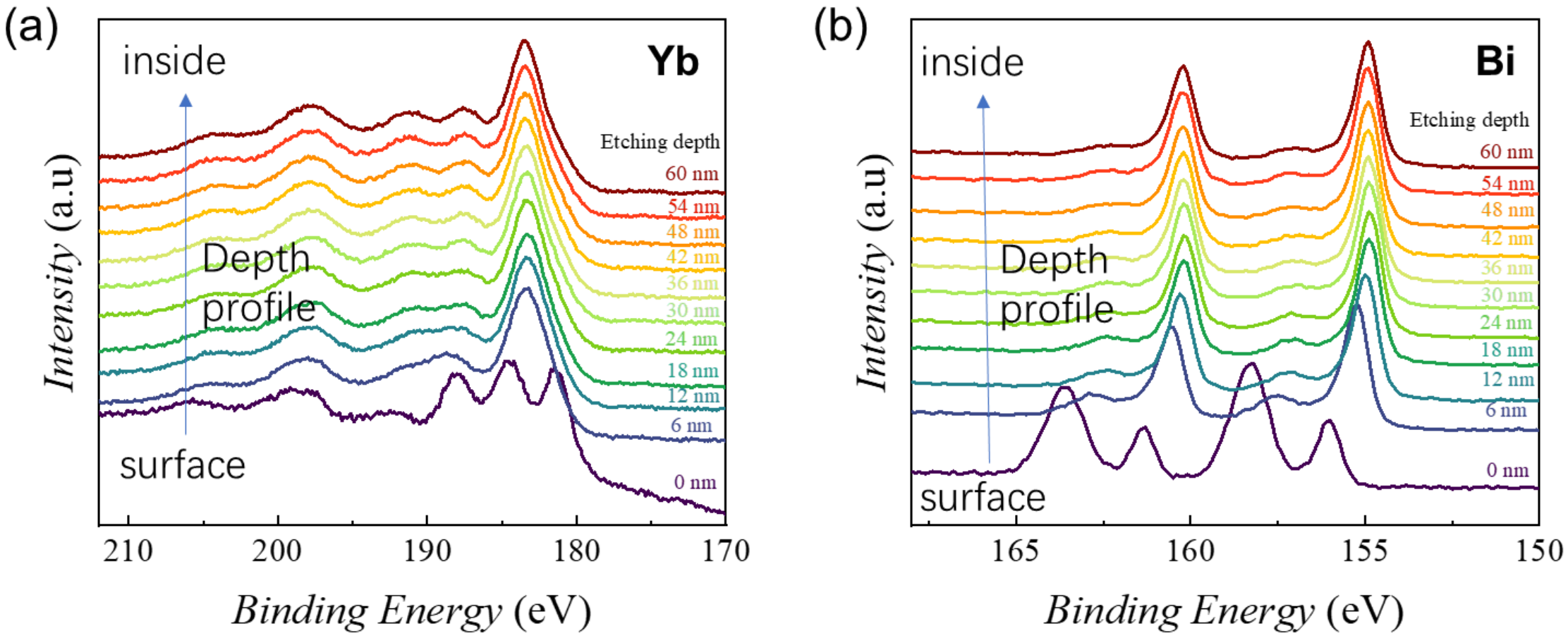

**Extended Data Fig. 11 | X-ray photoelectron spectroscopy measurement on $YbMnBi_2$.** Detection of (a) Yb and (b) Bi. Bi oxidization peaks are observed on the surface, but none of any oxide peak shows up in the inside even for an etching depth of only 6 nm.

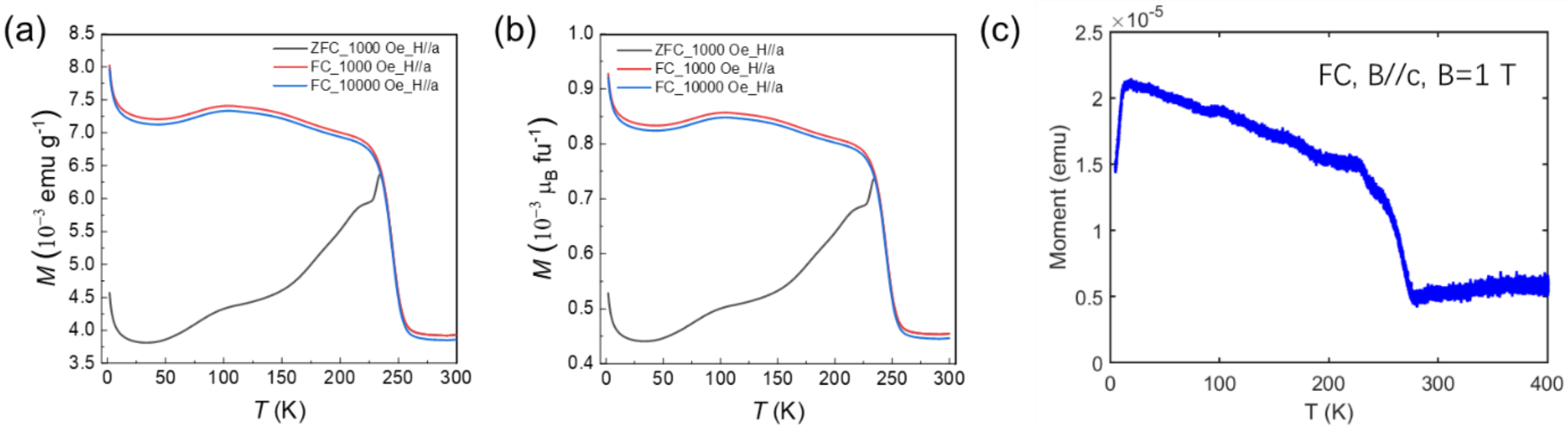

**Extended Data Fig. 12 Temperature dependence of the magnetization of $YbMnBi_2$.** (a) and (b) Magnetic field is applied in the ab-plane, and (c) Magnetic field is applied along the *c*-axis.

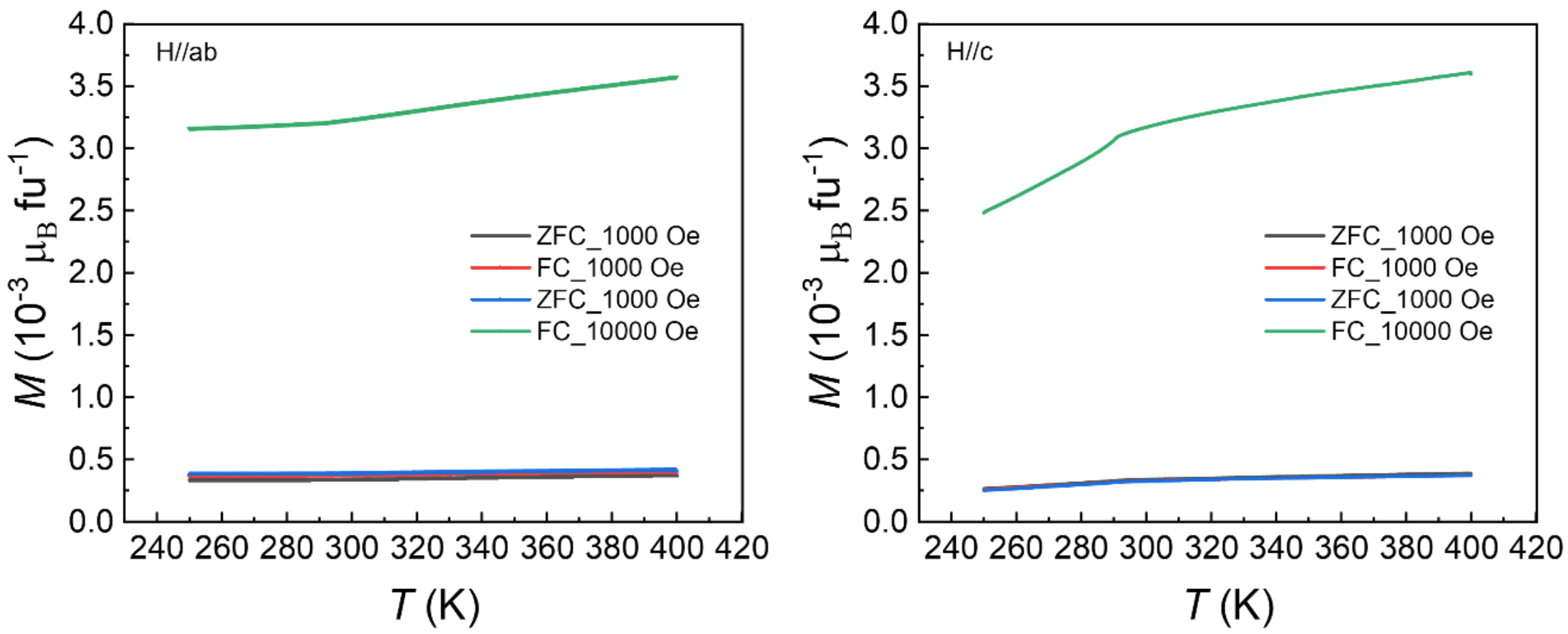

**Extended Data Fig. 13 Temperature dependence of the magnetization of $YbMnBi_2$ for in-plane and c-axis aligned magnetic fields.** We note that the experimental magnetization of $YbMnBi_2$ increases with increasing temperature over 250-400 K for both field directions. This behavior is inconsistent with the Curie expectation for localized moments and indicates that the bulk magnetic response is not governed by independent or weakly interacting local moments.

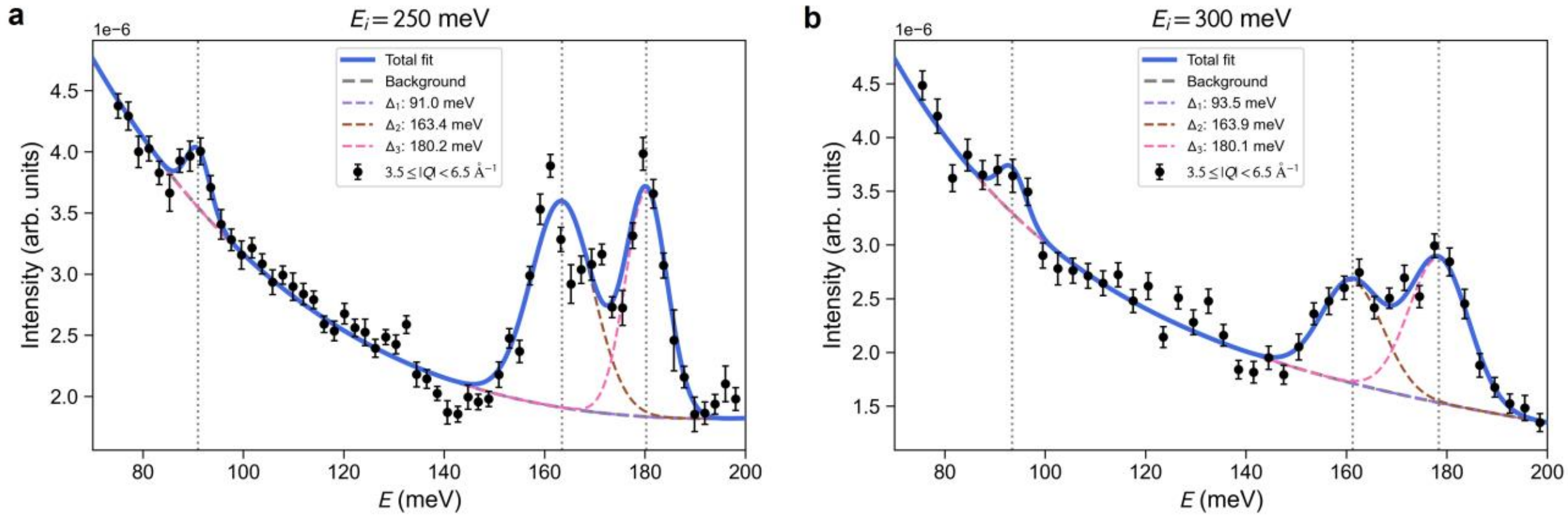

**Extended Data Fig. 14 Extraction of the $Yb^{3+}$ crystal-electric-field excitation energies and relative intensities from $E_i$ = 250 meV and 300 meV datasets.** Constant-$|Q|$ cuts integrated over $3.5 \leq |Q| < 6.5$ Å$^{-1}$ for (**a**) $E_i$ = 250 meV and (**b**) $E_i$ = 300 meV. Black symbols show the

experimental data. The solid blue curves are the total fits, consisting of three Gaussian peaks and a smooth background comprising a linear term and an exponentially decaying term. The gray dashed curves show the background contributions, while the colored dashed curves show the individual Gaussian components. The vertical dotted lines mark the fitted peak centers. The extracted excitation energies are 91.0, 163.4, and 180.2 meV for $E_i$ = 250 meV dataset, and 93.5, 163.9, and 180.1 meV for $E_i$ = 300 meV dataset. The integrated Gaussian peak areas were used to determine the relative transition intensities employed in the CEF analysis.